\documentclass[aps, twocolumn, nofootinbib,superscriptaddress]{revtex4-1}
\usepackage{amsmath}
\usepackage{amsfonts}
\usepackage{amssymb}
\usepackage{bbold}
\usepackage{ragged2e}
\usepackage[font=small,labelfont=bf,justification=justified]{caption}
\usepackage{hyperref}

\usepackage{color}
\newcommand{\rev}[1]{\textcolor{black}{#1}}
\newcommand{\x}{\times}

  \makeatletter
    \renewcommand\@make@capt@title[2]{%
     \@ifx@empty\float@link{\@firstofone}{\expandafter\href\expandafter{\float@link}}%
      {\textbf{#1}}\@caption@fignum@sep#2\quad}%
    \makeatother
   
\makeatletter 
\renewcommand{\fnum@figure}{\textbf{Figure~\thefigure}}
\makeatother

\usepackage{amsmath}
\usepackage{amsfonts}
\usepackage{graphicx}
\usepackage{verbatim} 
\usepackage{subfig}

\newcommand{\mc}{\mathcal}
\newcommand{\real}{\mathbb{R}}

\newcommand{\map}[3]{#1: #2 \rightarrow #3}
\newcommand{\transpose}{\mathsf{T}}

\begin{document}

\title{The control of brain network dynamics across diverse scales of space and time}

\author{Evelyn Tang}
\affiliation{Department of Bioengineering, School of Engineering \& Applied Science, University of Pennsylvania, PA 19104 USA}
\affiliation{Max Planck Institute for Dynamics and Self-Organization, G\"{o}ttingen 37079, Germany}
\author{Harang Ju}
\affiliation{Department of Bioengineering, School of Engineering \& Applied Science, University of Pennsylvania, PA 19104 USA}
\affiliation{Neuroscience Graduate Program, Perelman School of Medicine, University of Pennsylvania, PA 19104 USA}
\author{Graham L. Baum}
\affiliation{Department of Bioengineering, School of Engineering \& Applied Science, University of Pennsylvania, PA 19104 USA}
\affiliation{Neuroscience Graduate Program, Perelman School of Medicine, University of Pennsylvania, PA 19104 USA}
\author{David R. Roalf}
\affiliation{Department of Psychiatry, Perelman School of Medicine, University of Pennsylvania, PA 19104 USA}
\author{Theodore D. Satterthwaite}
\affiliation{Department of Psychiatry, Perelman School of Medicine, University of Pennsylvania, PA 19104 USA}
\author{Fabio Pasqualetti}
\affiliation{Department of Mechanical Engineering, University of California, Riverside, Riverside, CA 92521 USA}
\author{Danielle S. Bassett}
\affiliation{Department of Bioengineering, School of Engineering \& Applied Science, University of Pennsylvania, PA 19104 USA}
\affiliation{Department of Psychiatry, Perelman School of Medicine, University of Pennsylvania, PA 19104 USA}
\affiliation{Department of Physics \& Astronomy, College of Arts \& Sciences, University of Pennsylvania, PA 19104 USA}
\affiliation{Department of Electrical \& Systems Engineering, School of Engineering \& Applied Science, University of Pennsylvania, PA 19104 USA}
\affiliation{Department of Neurology, Perelman School of Medicine, University of Pennsylvania, PA 19104 USA}

\begin{abstract}
The human brain is composed of distinct regions that are each associated with particular functions and distinct propensities for the control of neural dynamics. However, the relation between these functions and control profiles is poorly understood, as is the variation in this relation across diverse scales of space and time. Here we probe the relation between control and dynamics in brain networks constructed from diffusion tensor imaging data in a large community based sample of young adults. Specifically, we probe the control properties of each brain region and investigate their relationship with dynamics across various spatial scales using the Laplacian eigenspectrum. In addition, through analysis of regional modal controllability and partitioning of modes, we determine whether the associated dynamics are fast or slow, as well as whether they are alternating or monotone. We find that brain regions that facilitate the control of energetically easy transitions are associated with activity on short length scales and slow time scales. Conversely, brain regions that facilitate control of difficult transitions are associated with activity on long length scales and fast time scales. Built on linear dynamical models, our results offer parsimonious explanations for the activity propagation and network control profiles supported by regions of differing neuroanatomical structure.
\end{abstract}

\maketitle
The brain is an inherently networked system that displays incredibly rich and complex dynamics \cite{bassett2017network,breakspear2017dynamic}. Building accurate models of those dynamics remains a key challenge that is fundamental to the field of neuroscience, with the potential to inform personalized medicine by predicting a patient's disease progression and response to therapy \cite{bansal2018personalized,proix2017individual,falcon2016new,ritter2015automated}. Efforts to build such models necessarily depend on the development of interdisciplinary approaches informed by dynamical systems theory, statistical physics, and network science as well as substantial knowledge of the intricacies of the underlying biology \cite{bassett2018on}. Because of the multiscale nature of the system \cite{breakspear2005dynamics,betzel2017multi}, there are some regimes of function or modalities of measurement whose dynamics are well-fit by one mathematical model and others whose dynamics are well-fit by another mathematical model \cite{friston2005models,friston2008hierarchical,deco2008dynamic,breakspear2017dynamic}. On the one hand, large-scale measurements of brain activity can display a wave-like nature and other characteristics consistent with linear, diffusive dynamics instantiated on a physically embedded system \cite{Honey2009,Galan2008,Raj2012,10.1371/journal.pcbi.1002435,Viventi2011}. On the other hand, fine time scale measurements of brain activity can display oscillatory characteristics consistent with nonlinear, synchronization dynamics thought to support distributed communication and computation \cite{kopell2014beyond,fries2015rhythms,palmigiano2017flexible,kirst2016dynamic}.

Evidence for the first type of dynamics comes, for example, from studies of both large- and small-scale neuronal processes whose intrinsic or endogeneous activity can be partially explained by simple linear models \cite{Honey2009,Galan2008}. Moreover, wave-like activity has been observed using functional magentic resonance imaging (fMRI) in the visual cortex, where traveling wave responses to visual stimuli are observed to propagate out from the fovea \cite{10.1371/journal.pcbi.1002435}. Notably, such traveling waves with various timescales have been observed across different regions of the brain \cite{Muller2018}, and it has been posited that the direction and wavelength of the wave could encode information transmitted between regions \cite{10.1371/journal.pcbi.1003260}. In addition to the outward propagation of activity, spiral waves also occur frequently in the neocortex \textit{in vivo}, both during pharmacologically induced oscillations and during sleep-like states \cite{HUANG2010978}, while seizures may manifest as recurrent spiral waves that propagate in the neocortex \cite{Viventi2011}. Such multiplicity of phenomena motivates the use of dynamical models that emphasize the spatially embedded nature of brain networks \cite{stiso2018spatial}, and the extended versus transient responses that embedding can support \cite{CHAUDHURI2015419}. An example is the use of finite element modeling to simulate the propagation of brain stimulation \cite{1741-2552-13-3-036022,6346881}, where effects are distributed on a localized part of brain tissue \cite{6347243}. Another example is that of diffusive models, where network diffusion over long time scales has been used to model disease progression in the brain for dementia \cite{Raj2012}, or to predict longitudinal patterns of atrophy and metabolism in Alzheimer's disease \cite{Raj2015}. 

Evidence for the second type of dynamics comes from studies of both large- and small-scale ensembles that produce rhythmic or oscillatory activity \cite{kopell2014beyond,fries2015rhythms,jones2016when}. For example, scalp electrodes in humans can be used to measure rhythms of certain frequencies, which in turn have been associated with different cognitive processes and behavioral responses \cite{palmigiano2017flexible,kirst2016dynamic}. For example, attentional control via the inhibition of behaviorally irrelevant stimuli and motor responses has been associated with the synchronization of $\alpha$ and $\beta$ rhythms between right inferior frontal and primary sensory neocortex \cite{sacchet2015attention}. Similarly, higher frequency $\gamma$ rhythms in local synchronization have been observed during visual responses, while lower frequency $\beta$ rhythms reflecting coherence over parietal and temporal cortices have been observed during activities that required more multi-modal sensory integration \cite{doi:10.1093cercor9.2.137}. Initial efforts sought to explain such phenomena using a simple neuronal model with conduction delays, where excitatory units describe pyramidal cells and inhibitory units describe interneurons \cite{Kopell1867}. More recent work suggests that the $\beta$ rhythms specifically can emerge from the integration of nearly synchronous bursts of excitatory synaptic drive targeting proximal and distal dendrites of pyramidal neurons, where the defining feature of a $\beta$ event was a strong distal drive to supragranular and infragranular layers of cortex that lasted one $\beta$ period \cite{sherman2016neural}. Notably, the laminar architecture of the neocortical network has also proven critical for explaining $\gamma$ rhythms \cite{lee2013distinguishing}.

\begin{figure*}
\includegraphics[width=0.7\linewidth]{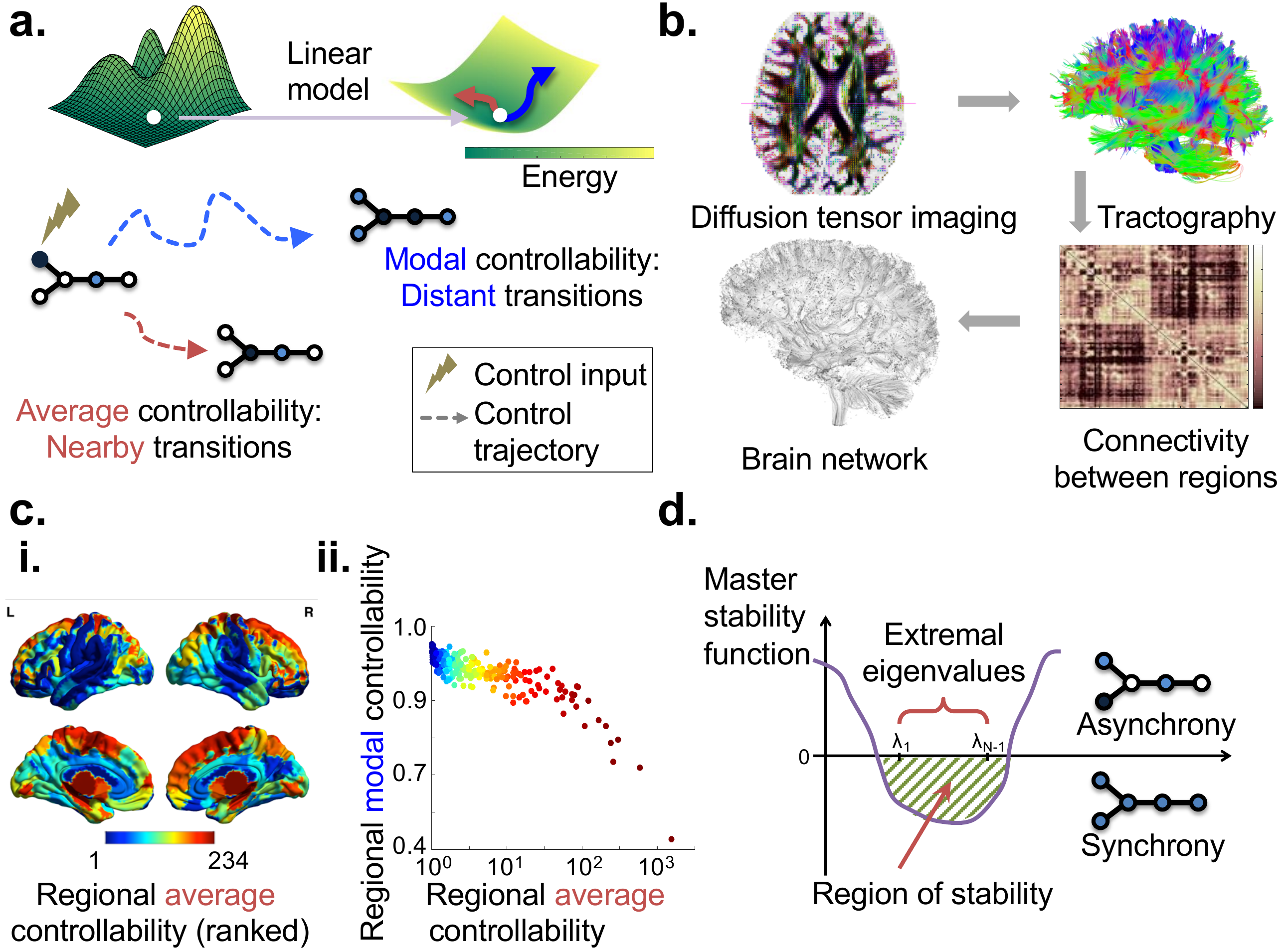}
\caption{\textbf{Controllability and synchronizability in brain networks. (a)} While the brain displays nonlinear dynamics (top left), linear models (top right) have shown great utility in predicting such dynamics across spatial scales \cite{Honey2009,Galan2008,muldoon2016stimulation}. We study two such metrics: Average controllability provides an intuitive measure of the structural support for moving the brain to easy-to-reach states (short red transition), whereas modal controllability provides an intuitive measure of the structural support for moving the brain to difficult-to-reach states (long blue transition). \textbf{(b)} Diffusion tensor imaging measures the direction of water diffusion in the brain. From this data, white matter streamlines can be reconstructed that connect brain regions in a structural network. \textbf{(c.i)} Regional average controllability ranked on $N=234$ brain regions of a group-averaged network for visualization purposes. \textbf{(c.ii)} Regions with high average controllability tend to display low modal controllability: $\rho=-0.76$, $df=233$, $p<1\times10^{-16}$. \textbf{(d)} We operationalize a synchronous state as a state in which all nodes have the same activity magnitude. Such a state is stable when the master stability function is negative for all positive eigenvalues of the graph Laplacian \cite{PhysRevLett.80.2109}. Following \cite{tang2017developmental}, we use the inverse spread of the Laplacian eigenvalues $1/\sigma^2(\{\lambda_{i}\})$ as a measure of global synchronizability. Adapted with permission from \cite{tang2017developmental}.} \label{fig:schematic}
\end{figure*}

Although empirical studies find pervasive evidence of these two distinct types of dynamics, there is little theoretical framework to model how the spatiotemporal characteristics of these dynamics can inform the control properties of specific brain regions, which are interconnected by a fixed structural network. To address this question, recent efforts have begun to study how such distinct dynamics can be guided or controlled by local energy input \cite{tang2018control}. For the control of those features of the dynamics that are well-approximated by linear systems theory, efforts have stipulated a linear model of the dynamics dependent upon the structural network implicit in white matter connectivity \cite{gu2015controllability,taylor2015optimal}. This framework of network control has been useful to understand brain anatomy and regional function across multiple species including the nematode \emph{C. elegans} \cite{Yan2017}, the fly \emph{Drosophila} \cite{kim2018role}, the mouse \cite{kim2018role}, the macaque \cite{gu2015controllability}, and the human \cite{tang2017developmental}. Network control theory has also been used to explore the control profiles of brain regions based on their network connectivity \cite{FP-SZ-FB:13q}, and hence to map control propensities on to canonically understood cognitive functions \cite{gu2015controllability,tang2017developmental,cornblath2018sex} and their alterations in psychiatric disorders \cite{jeganathan2018fronto}. For the control of those features of the dynamics that require nonlinear systems theory, network control is more difficult \cite{motter2015networkcontrology,muldoon2016stimulation}, but some initial work has begun to assess the utility of control principles for seizure abatement \cite{taylor2015optimal} and to determine routes along which information can be propagated by transient synchrony \cite{palmigiano2017flexible,kirst2016dynamic}. In extending these ongoing efforts, a key challenge remains to build models that can predict not only the generic control properties of a system, but also the length scales over which control-induced dynamics will propagate, and to link these length scales to the time scales of dynamical transitions or control of activity enacted by regional drivers.

Here, we perform initial numerical experiments to address this challenge by relating new developments in the linear control of regional activity in brain networks and their predicted time scales, with the length scales implicit in nonlinear dynamical models of inter-regional synchronization. We use the framework of network control \cite{RevModPhys.88.035006} to probe the relation between control and dynamics in brain networks constructed from diffusion tensor imaging data in a young adult subset of the Philadelphia Neurodevelopmental Cohort, a large community based sample of youth \cite{satterthwaite2016philadelphia,calkins2015philadelphia}. Use of this large sample allows us to probe relations between control and dynamics in an ensemble of brain networks in which control propensities vary extensively \cite{tang2017developmental}. The paper is organized as follows. Following a description of the formalism in Sec.~\ref{sec:framework}, we turn in Sec.~\ref{sec:control} to a review of network controllability metrics and their predictions regarding the control profiles of certain brain regions (see the Appendix for details regarding brain network construction from neuroimaging data). In Sec.~\ref{sec:scales}, we consider metrics for the characterization of spatial scales of synchronization in network dynamics \cite{Atasoy2016} and for the control of such synchronization \cite{PhysRevE.90.012909}. In Sec.~\ref{sec:time}, we consider metrics for the characterization of temporal scales of control in network dynamics, specifically assessing the role of modal controllers in driving transient versus extended temporal modes of brain activity \cite{FP-SZ-FB:13q}. In Sec.~\ref{sec:topology}, we compare the relation between control of time scales and control of length scales that we observe in brain networks to those that we observe in networks constructed via the rules of preferential attachment, gaining insight into the dependence of spatiotemporal control on the underlying topology. In Sec.~\ref{sec:discussion}, we place our findings within the broader context of empirical results from neuroimaging experiments. We also discuss the implications of our findings for our understanding of how various controllers are associated with distinct types of dynamics, and in turn how distinct dynamics are associated with different regional controllers. We close with a brief discussion of potential future directions.

\section{Mathematical Framework}
\label{sec:framework}

In this section, we describe a formalism for building network models of brain architecture and states, and for studying network controllability of brain dynamics. 

\subsection{Network models of brain architecture and states}

Here, we study 190 brain networks describing the white matter connectivity of youth in the Philadelphia Neurodevelopmental Cohort. For each network composed of $N=234$ cortical and subcortical brain regions, $\mathbf{A} \in \real^{N \times N}$ is a symmetric and weighted adjacency matrix whose elements indicate the number of white matter streamlines connecting two different brain regions indexed with $(i,j)$. For further details regarding the construction of brain networks from neuroimaging data see the Appendix, and for prior work in this data set see \cite{tang2017developmental,baum2017modular,baum2018impact,cornblath2018sex}. 

More formally, a networked system such as the brain can be represented by the graph $\mc G = (\mc V, \mc E)$, where $\mc V$ and $\mc E$ are the vertex and edge sets, respectively. Let $a_{ij}$ be the weight associated with the edge $(i,j) \in \mc E$, and define the \emph{weighted adjacency matrix} of $\mc G$ as $A = [a_{ij}]$, where $a_{ij} = 0$ whenever $(i,j) \not\in \mc E$. We associate a real value (\emph{state}) with each node, collect the node states into a vector (\emph{network state}), and define the map $\map{x}{\mathbb{N}_{\ge 0}}{\mathbb{R}^n}$ to describe the evolution (\emph{network dynamics}) of the network state over time. 

Next we seek to define an appropriate equation of state. Based on prior work demonstrating the utility of simple linear models in predicting intrinsic brain dynamics across spatial scales \cite{Honey2009,Galan2008}, we employ a simplified noise-free linear discrete-time and time-invariant network model \cite{gu2015controllability}: \begin{equation}\label{eq: linear network}
  \mathbf{x} (t+1) = \mathbf{A} \mathbf{x}(t) + \mathbf{B}_{\mc K} \mathbf{u}_{\mc K} (t) ,
\end{equation}
where $\map{\mathbf{x}}{\real_{\ge 0}}{\real^N}$ describes the state (i.e., a measure of the electrical charge, oxygen level, or firing rate) of brain regions over time, and $\mathbf{A} \in \real^{N \times N}$ is the structural connectivity described in the previous section. Hence the size of the vector $\mathbf{x}$ is given by $N$, and the value of $\mathbf{x}$ describes the brain activity of that region.

\begin{figure*}
\includegraphics[width=0.7\linewidth]{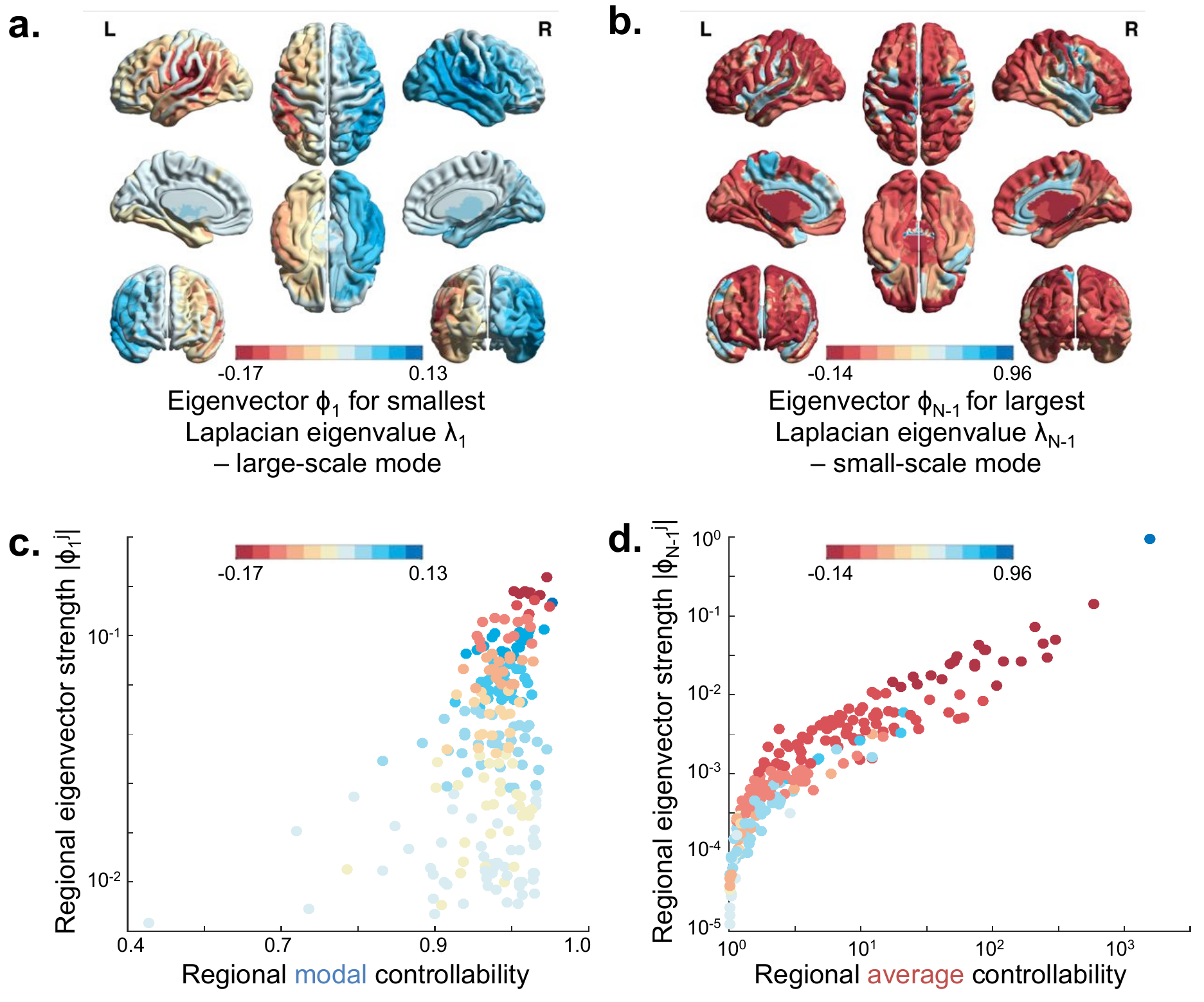}
\caption{\textbf{Synchronizability and the spatial extent of predicted harmonic waves. (a)} Spatial distribution of the eigenvector $\phi_{1}$ for the smallest Laplacian eigenvalue $\lambda_{1}$, showing which regions on a group-averaged brain network most strongly contribute to this large-scale mode. \textbf{(b)} Spatial distribution of the eigenvector $\phi_{N-1}$ for the largest Laplacian eigenvalue $\lambda_{N-1}$, showing which regions most strongly contribute to this small-scale mode. \textbf{(c)} Regions most relevant for this large-scale mode $|\phi_{1}^j|$ are positively correlated with regions of high modal controllability: $\rho=0.27$, $df=233$, $p<1\times10^{-4}$. \textbf{(d)} Regions most relevant for this small-scale mode $|\phi_{N-1}^j|$ are positively correlated with regions of high average controllability: $\rho=0.95$, $df=233$, $p<1\times10^{-5}$. Note that the color in panels \textbf{(a)-(b)} is recapitulated by values along the vertical axis in panels \textbf{(c)-(d)}, simply for ease of identifying relevant brain regions with the given results.}\label{fig:spatial}
\end{figure*}

\subsection{Network controllability of brain dynamics}

To assess controllability of this networked system, we must first ensure its stability and define input points for the injection of control energy. We note first that the diagonal elements of the matrix $\mathbf{A}$ satisfy $A_{ii}=0$. Next we note that to assure Schur stability, we divide the matrix by $1+\xi_0(\mathbf{A})$, where $\xi_0(\mathbf{A})$ is the largest eigenvalue of $\mathbf{A}$. The input matrix $\mathbf{B}_{\mc K}$ identifies the control points $\mc K$ in the brain, where $\mc K = \{k_1, \dots, k_m \}$ and \begin{align}\label{eq: B}
  B_{\mc K} =
  \begin{bmatrix}
    e_{k_1} & \cdots & e_{k_m}
  \end{bmatrix},
\end{align}
and $e_i$ denotes the $i$-th canonical vector of dimension $N$. The input $\mathbf{u}_{\mc K}: \mathbb{N} \rightarrow
  \mathbb{R}^{|\mc{K}|}$ denotes the control strategy, i.e. the
  control signal injected into the network via the nodes
  $\mc{K}$. More precisely, it is a map from the set of natural
  numbers (representing the time instant), and the set of real
 vectors of length $|\mc{K}|$ (representing the actual
  control vector).

We can now study the \emph{controllability} of this dynamical system, which refers to the possibility of driving the state of the system to a specific target state by means of an external control input \cite{rek-ych-skn:63_2}. Classic results in control theory ensure that controllability of the network \eqref{eq: linear network} from the set of nodes $\mc K$ is equivalent to the controllability Gramian $\mathbf{W}_{\mc K}$ being invertible, where 
\begin{equation}
  \mathbf{W}_{\mathcal{K}} = \sum_{\tau =0}^{\infty}\mathbf{A}^\tau
  \mathbf{B}_{\mathcal{K}}\mathbf{B}_{\mathcal{K}}^\transpose \mathbf{A}^\tau .
\end{equation}
Consistent with \cite{gu2015controllability,tang2017developmental,cornblath2018sex,jeganathan2018fronto}, we use this framework to choose control nodes one at a time, and thus the input matrix $\mathbf{B}_{\mc K}$ in fact reduces to a one-dimensional vector, e.g., $\mathbf{B}_{\mc K}=\begin{pmatrix}1& 0& 0& ... \end{pmatrix}^T$ when the first brain region is the control node. In this case, $\mc K$ simply describes this control node, i.e. the controllability Gramian can be indexed by the $i$-th control node that it describes: $\mathbf{W}_{i}$.

\section{Controllability metrics for brain networks}
\label{sec:control}

Within the network controllability framework, we study two different control strategies that describe the ability to move the network into different states defined as patterns of regional activity. While the brain certainly displays non-linear activity, we study a linear model of this more complex nonlinear process which is valid in a neighborhood of the linearization point (see Fig. \ref{fig:schematic}a). Modeling of brain activity in large-scale regional networks shows that the linear approximation provides fair explanatory power of resting state fMRI BOLD data \cite{Honey2009}. Further, studies of this controllability framework using non-linear oscillators connected with coupling constants estimated from white matter tracts shows a good overlap with the linear approximation \cite{muldoon2016stimulation}.

The energy landscape  of the linear model is quadratic, with at most one energy minimum (at the origin), corresponding to the only state that can be  maintained without cost. Note that this origin of the linear system corresponds to the linearization point, i.e. the equilibrium configuration of the nonlinear model from which the linear model is obtained. Using the linear model described in the previous section, we can introduce two controllability metrics.
Intuitively, average controllability describes the ease of transition to many states nearby on an energy landscape, while modal controllability describes the ease of transition to a state distant on this landscape (see Fig. \ref{fig:schematic}a).

\emph{Average controllability} of a network equals the average input energy from a set of control nodes and over all possible target states. As a known result, average input energy is proportional to $\text{Trace}( \mathbf{W}_{\mathcal{K}} ^{-1})$, the trace of the inverse of the controllability Gramian. Instead and consistent with \cite{gu2015controllability,tang2017developmental,cornblath2018sex}, we adopt $\text{Trace}( \mathbf{W}_{\mathcal{K}} )$ as a measure of average controllability for two main reasons: first, $\text{Trace}( \mathbf{W}_{\mathcal{K}} ^{-1})$ and $\text{Trace}( \mathbf{W}_{\mathcal{K}} )$ satisfy a relation of inverse proportionality, so that the information obtained from the two metrics are correlated with one another and, second, $ \mathbf{W}_{\mathcal{K}} $ is typically very ill-conditioned even for coarse network resolutions, so that $\text{Trace}( \mathbf{W}_{\mathcal{K}} ^{-1})$ cannot be accurately computed even for small brain networks. It should be noted that $\text{Trace}( \mathbf{W}_{\mathcal{K}})$ encodes a well-defined control metric, namely the energy of the network impulse response or, equivalently, the network $H_2$ norm \cite{kailath1980linear}. As discussed above, when a brain region $i$ serves as a control node, the resulting Gramian can be indexed as $\mathbf{W}_{i}$, in order to compute the regional average controllability.

To calculate the trace of the Gramian, we avoid the numerical difficulties associated with solving the
  Lyapunov equation \cite{hewer1988sensitivity,gahinet1990sensitivity,hammarling1982numerical,sorensen2002bounds}. Given the stability of our matrix $\mathbf{A}$, we can use the following: 
  \begin{align*}
    \text{Trace}( \mathbf{W}_{\mathcal{K}}) &= \text{Trace}\left(\sum_{\tau =0}^{\infty}\mathbf{A}^\tau
  \mathbf{B}_{\mathcal{K}}\mathbf{B}_{\mathcal{K}}^\transpose \mathbf{A}^\tau\right) \\
  &=\sum_{\tau =0}^{\infty} \text{Trace}\left(\mathbf{A}^\tau
  \mathbf{B}_{\mathcal{K}}\mathbf{B}_{\mathcal{K}}^\transpose \mathbf{A}^\tau\right) \\
  &=\sum_{\tau =0}^{\infty} \text{Trace}\left(\mathbf{A}^{2\tau}
  \mathbf{B}_{\mathcal{K}}\mathbf{B}_{\mathcal{K}}^\transpose\right) \\
  &=\left(\sum_{\tau =0}^{\infty} \mathbf{A}^{2\tau}\right)
  \left(\mathbf{B}_{\mathcal{K}}\mathbf{B}_{\mathcal{K}}^\transpose\right) \\
  &=\left(\mathbf{I}-\mathbf{A}^{2\tau}\right)^{-1}
  \left(\mathbf{B}_{\mathcal{K}}\mathbf{B}_{\mathcal{K}}^\transpose\right).
  \end{align*}
Here we used some properties of our system: that ${\mc K}$ contains only one node and $\mathbf{A}$ is symmetric.

\emph{Modal controllability} refers to the ability of a node to control each time-evolving mode of a dynamical network, and can be used to identify states that are difficult to control from a set of control nodes. Modal controllability is computed from the eigenvector matrix $V = [v_{ij}]$ of the network adjacency matrix $\mathbf{A}$. By extension from the PBH test \cite{kailath1980linear}, if the entry $v_{ij}$ is small, then the $j$-th mode is poorly controllable from node $i$. Following \cite{pasqualetti2014controllability}, we define \begin{equation}
\phi_i =  \sum_{j} (1 - \xi_j^2 (A)) v_{ij}^2,\label{eq:modal}
\end{equation}
as a scaled measure of the controllability of all $N$ modes $\xi_0 (\mathbf{A}),\dots, \xi_{N-1} (\mathbf{A})$ from the brain region $i$ -- allowing the computation of regional modal controllability. Regions with high modal controllability are able to control all of the dynamic modes of the network, and hence to drive the dynamics towards hard-to-reach configurations.

We study these metrics in the 190 structural brain networks derived from diffusion tensor imaging data (see Fig. \ref{fig:schematic}b). On a group-representative brain network constructed by averaging all 190 subject-specific networks, we calculate these controllability metrics and show their distribution across the brain (see Fig. \ref{fig:schematic}c.i). Consistent with prior work in both this and other data sets \cite{gu2015controllability,tang2017developmental}, we find that regions with high average controllability tend to be located in network hubs associated with the default-mode system, a set of regions that tend to be active during intrinsic processing. Interestingly and again consistent with prior work in this and other data sets \cite{gu2015controllability,tang2017developmental,jeganathan2018fronto,wuyan2018benchmarking}, regions with high modal controllability tend to display low average controllability (see Fig. \ref{fig:schematic}c.ii), and are predominantly found in regions of the brain that become active during tasks that demand high levels of executive function or cognitive control.

\section{Controllability of spatial scales: Network dynamics in harmonic waves}
\label{sec:scales}

In contrast to the time-varying or transient dynamics that can be described using models mentioned in the previous section, recent observations of brain dynamics also includes sustained patterns of collective activity across relatively large areas of the brain \cite{Muller2018,Roberts2019}. Such wave patterns instead call for a different description, one better suited by a framework developed for networked oscillators and synchronized activity.

Specifically, the previous linear model (Eq. 1) is not relevant to wave-like dynamics, even as it is supported by other empirical observations \cite{Honey2009,Galan2008,Yan2017}. Therefore, to complement our previous efforts to study the control of brain state transitions via a simple linear model of network dynamics, here we turn to the question of how to control the synchrony of harmonic waves \cite{Atasoy2016} which naturally arise in the same underlying white matter architecture. Notably, extensive prior work in the dynamical systems literature has considered ways in which to measure the synchronizability of a networked system. To be clear, here we define the synchronizability as a measure of the ability of a network to persist in a single synchronous state $\mathbf{s} (t)$, i.e. $\mathbf{x}_1 (t) = ... = \mathbf{x}_n (t+1) = \mathbf{s} (t)$. The master stability function (MSF) allows analysis of the stability of this synchronous state without detailed specification of the properties of the dynamical units (see Fig. \ref{fig:schematic}d; \cite{PhysRevLett.80.2109}). Within this framework, linear stability depends upon the positive eigenvalues  $\{\lambda_{i}\}, i=1, ... ,N-1$ of the Laplacian matrix $L$ defined by $L_{ij}=\delta_{ij}\sum_{k}A_{ik}-A_{ij}$. The synchronous state is stable when the master stability function is negative for all positive eigenvalues of the graph Laplacian (see Fig. \ref{fig:schematic}d).

For the purposes of control, it is useful to note that the structure of the graph Laplacian provides information regarding the most likely modes of standing waves: the vibrational states of a dynamical system in which the frequency of vibration is the same for all elements. These modes are given by the eigenvectors of the graph Laplacian. For example, the eigenvector $\phi_{1}$ of the smallest positive Laplacian eigenvalue $\lambda_{1}$ is an odd mode associated with large-scale waves \cite{Atasoy2016,PhysRevLett.80.2109} (Fig.~\ref{fig:spatial}a). In contrast, the eigenvector $\phi_{N-1}$ of the largest Laplacian eigenvalue $\lambda_{N-1}$ is an even mode associated with small-scale waves (Fig.~\ref{fig:spatial}b). Recent developments in push-pull control capitalize on these relations to control the transition into and out of synchrony by targeting network nodes that have a large weight for a given extremal eigenvector of the Laplacian \cite{PhysRevE.90.012909}. 

\begin{figure}
\includegraphics[width=0.95\linewidth]{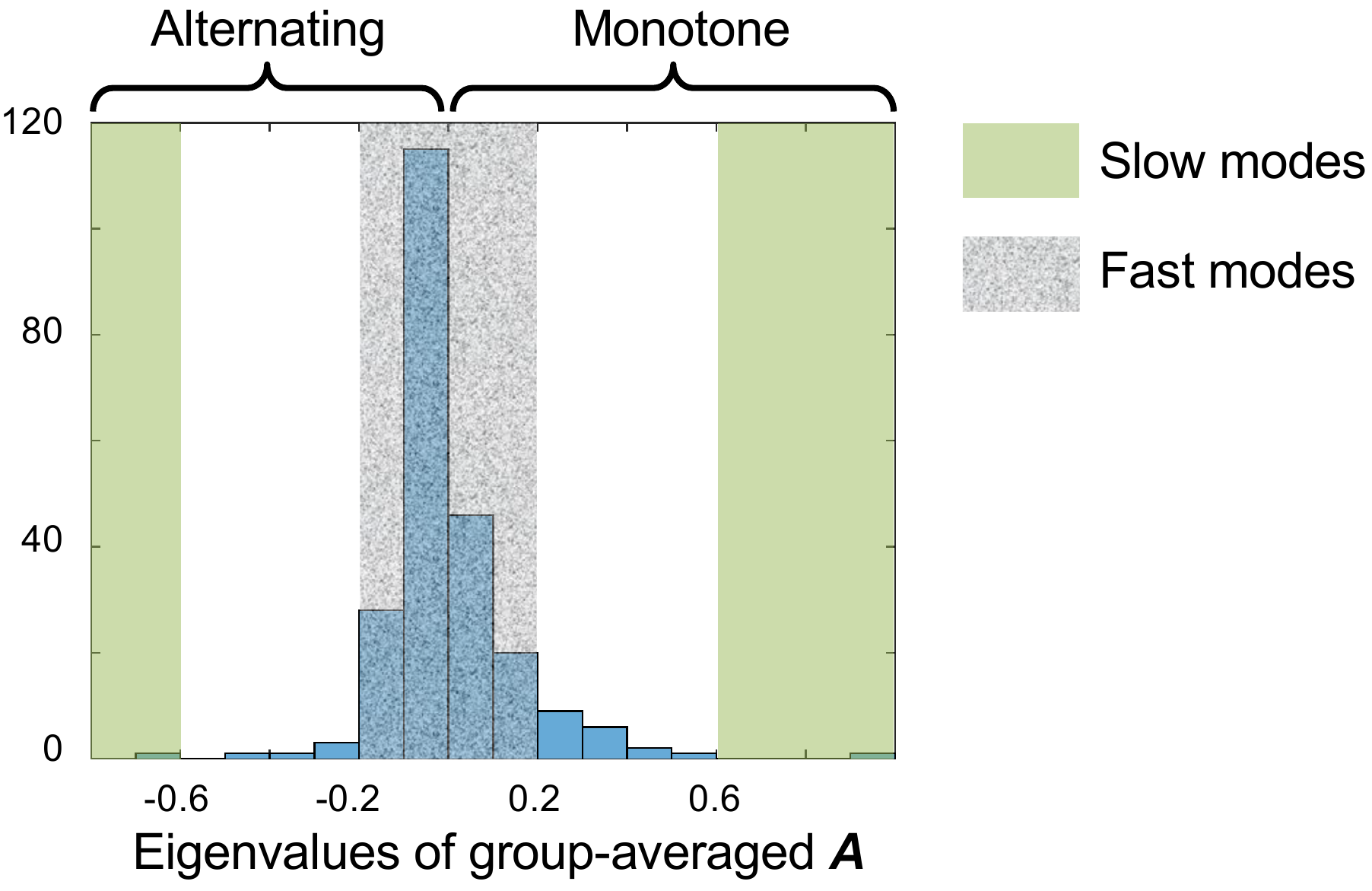}
\caption{\textbf{Extracting fast or slow, and alternating or monotone modes of network dynamics.} Here we show the histogram of $N=234$ eigenvalues of the group representative brain network, constructed by averaging all 190 subject-specific brain networks. Given the trimodal distribution of eigenvalues, we partition them into a densely populated cluster of $|\xi_j|<0.2$, a lightly populated cluster of $0.2<|\xi_j|<0.6$, and the remaining $|\xi_j|>0.6$ eigenvalues that are separated from the other modes with a clear gap. This partitioning can be done for both positive and negative eigenvalues, respectively, which correspond to the monotone and alternating modes of the system. From left to right, these groups are $\xi_j<-0.6$ (slow alternating), $-0.2<\xi_j<0$ (fast alternating), $0<\xi_j<0.2$ (fast monotone), and $\xi_j>0.6$ (slow monotone), respectively.} \label{fig:spectrum}
\end{figure}

\begin{figure*}
\includegraphics[width=0.7\linewidth]{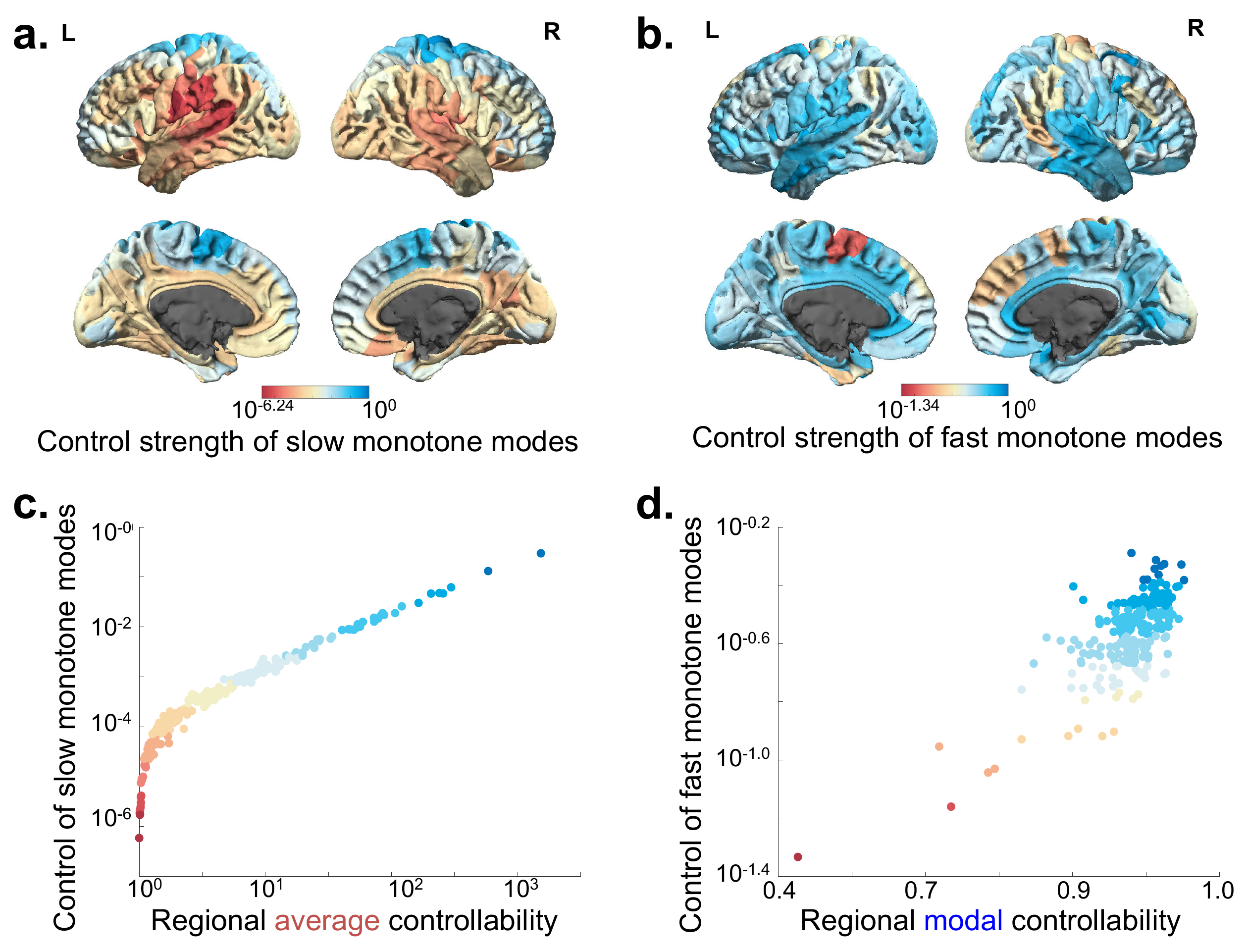}
\caption{\textbf{Monotone modes and their control profiles across the brain. (a)} Spatial distribution of the controllability of slow monotone modes ($\xi_j>0.6$), showing which regions on a group-representative brain network most strongly contribute. \textbf{(b)} Spatial distribution of the controllability of fast monotone modes ($0<\xi_j<0.2$), showing which regions on a group-representative network most strongly contribute. \textbf{(c)} Regions most relevant for control of slow monotone modes tended to be regions of high average controllability: $\rho=0.99$, $df=233$, $p<1\times10^{-4}$. \textbf{(d)} Regions most relevant for control of fast monotone modes tended to be regions of high modal controllability: $\rho=0.59$, $df=233$, $p<1\times10^{-4}$. Note that the color in panels \textbf{(a)-(b)} is recapitulated by values along the vertical axis in panels \textbf{(c)-(d)}, simply for ease of identifying relevant brain regions with the given results.} \label{fig:poscontrol}
\end{figure*}

To investigate the control of synchrony in the harmonic waves of brain networks, we calculate the eigenvectors of the Laplacian for each subject, and then we average the eigenvectors across all of the 190 subjects. We observe that the strength of control for harmonic waves of a network can be understood in terms of the network's predisposition to average and modal control. First, we observe that the regional strength of the large-scale waves $|\phi_{1}^j|$ is positively correlated with regional modal controllability (Spearman correlation coefficient $\rho=0.27$, $df=233$, $p<1\times10^{-4}$; Fig.~\ref{fig:spatial}c). This relation suggests that regions that enable synchronous behavior over long distances are also predicted to be effective in moving the brain to energetically distant states (see a list of the highest and lowest such regions in Table \ref{tab:lap1} of the Appendix). Second, we observe that the regional strength of the small-scale waves $|\phi_{N-1}^j|$ is positively correlated with regional average controllability (Spearman correlation coefficient $\rho=0.95$, $df=233$, $p<1\times10^{-5}$; Fig. \ref{fig:spatial}d). This relation suggests that regions that enable synchronous behavior over short distances are also predicted to be effective in moving the brain to energetically nearby states (see a list of the highest and lowest such regions in Table \ref{tab:lap234} of the Appendix). To confirm that our results do not depend on the stage at which we average over subjects, we also calculate these correlations for the Laplacian spectra obtained from a group-averaged network (procedure described in Section II). We find that the  regional strength of the large-scale waves in the group network is still positively correlated with regional modal controllability (Spearman correlation coefficient $\rho=0.28$, $df=233$, $p<1\times10^{-4}$), while the regional strength of the small-scale waves in the group network is still positively correlated with regional average controllability (Spearman correlation coefficient $\rho=0.86$, $df=233$, $p<1\times10^{-5}$). These results demonstrate that our findings do not depend unduly on the stage at which we average over subjects.

\begin{figure*}
\includegraphics[width=0.7\linewidth]{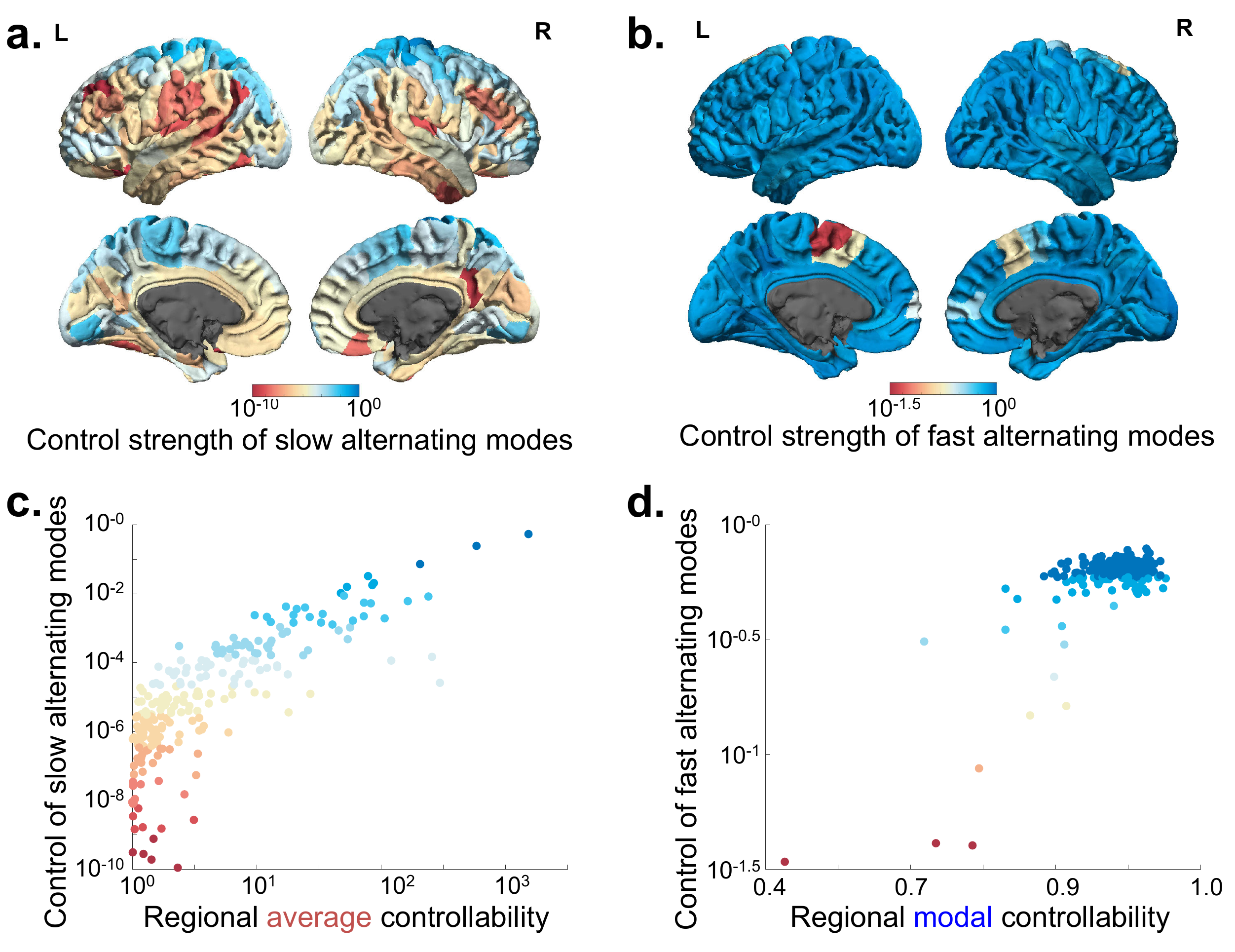}
\caption{\textbf{Alternating modes and their control profiles across the brain. (a)} Spatial distribution of the controllability of slow alternating modes ($\xi_j<-0.6$), showing which regions on a group-representative brain network most strongly contribute. \textbf{(b)} Spatial distribution of the controllability of fast alternating modes ($-0.2<\xi_j<0$), showing which regions on a group-representative network most strongly contribute. \textbf{(c)} Regions most relevant for the control of slow alternating modes tended to be regions of high average controllability: $\rho=0.83$, $df=233$, $p<1\times10^{-4}$. \textbf{(d)} Regions most relevant for the control of fast alternating modes tended to be regions of high modal controllability: $\rho=0.24$, $df=233$, $p<2\times10^{-4}$. Note that the color in panels \textbf{(a)-(b)} is recapitulated by values along the vertical axis in panels \textbf{(c)-(d)}, simply for ease of identifying relevant brain regions with the given results.} \label{fig:negcontrol}
\end{figure*}

\section{Controllability of temporal scales: Network dynamics in fast or slow, alternating or monotone modes}
\label{sec:time}

To complement the findings reported in the previous section explicating the spatial scales of control, we next turn to a consideration of the time scales of control. We note that in a discrete-time system, the eigenvalues of the network adjacency matrix $\mathbf{A}$ give the rate of decay of the various supported modes of activity. These modes of a linear time-invariant system are the
  elementary components of the dynamic response of the system to
  stimuli. Each mode gives rise to an elementary response (or trajectory),
  and the combination of the modes generates more complex responses and trajectories. Distinct processes and time scales of activity are observed in the brain, where rhythms of certain frequencies have been associated with different cognitive processes and behavioral responses \cite{palmigiano2017flexible,kirst2016dynamic}. For instance, higher frequency $\gamma$ rhythms in local synchronization emerge during visual responses, while lower frequency $\beta$ rhythms reflecting coherence over parietal and temporal cortices have been observed during activities that required more multi-modal sensory integration \cite{doi:10.1093cercor9.2.137,sherman2016neural}. In our discrete time linear model, the eigenvalues with large magnitude decay slowly while the eigenvalues with small magnitude decay quickly. The relationship between eigenvalues and the rate of decay of associated modes of activity suggests that the control of such modes can be analyzed separately for different time scales. While our prior expression for modal controllability (Eq.~\ref{eq:modal}) summed over all $N$ modes of the adjacency matrix $\mathbf{A}$, we can instead partition the sum into different groups based on the eigenvalues $\xi_j$ of the modes. For instance, a region $i$ can be evaluated for its controllability of fast modes with $\phi_i^{\textrm{fast}} = \sum_{j}^{\textrm{small } |\xi_j|} v_{ij}^2$ and for its controllability of slow modes with $\phi_i^{\textrm{slow}}= \sum_{j}^{\textrm{large } |\xi_j|} v_{ij}^2$. Note that in this case, all the eigenvalues have magnitude smaller than 1. Further, in the above expression we have also removed the scaling factor in Eq. \ref{eq:modal} of $\left(1 - \xi_j^2 (A)\right)$ as this factor is unnecessary for the calculation of control time-scales; in supporting calculations, we also verified that the inclusion or exclusion of this factor does not alter our subsequent findings.

To investigate regional propensities
  for the control of fast versus slow modes, we considered the
  group-representative network constructed by averaging all 190
  subject-specific brain networks. We examined the spectrum of this
  representative network (see Fig. 3), and note that there are
    three well-separated groups of eigenvalues. In the center, there
    is a densely populated cluster of magnitude between 0 and 0.2
    (fast), which have at least double the modes in each bin of width
    0.1 as compared to all the bins outside this cluster. On the two
    ends, there are sparse modes with magnitude above 0.6 (slow) that
    have a clear gap from all the other
    modes. In between these above mentioned groups, there is a lightly
    populated cluster of eigenvalue magnitude between 0.2 and 0.6,
    which are distinct from the other groups due to the reasons
    mentioned. Given this visually trimodal distribution, it is hence
    natural to partition the eigenvalues into three groups that differ
    in their relevant time scales: $|\xi_j|<0.2$ (fast),
    $0.2<|\xi_j|<0.6$ (medium), and $|\xi_j|>0.6$ (slow).

Now, it is also important to note that eigenvalues carry
  additional information: the sign of the eigenvalue determines the
  nature of the system's response, where a mode with a positive
  eigenvalue decays monotonically with each time step, while a mode
  with a negative eigenvalue has a decaying response that alternates
  between positive and negative values at each time step. To allow
  sensitivity to these monotone and alternating modes, we can restrict
  the range in the sum of $\phi_i^{\textrm{fast}}$ and
  $\phi_i^{\textrm{slow}}$ to the positive or to the negative
  eigenvalues. By doing so, we now have four groups of interest:
  $0<\xi_j<0.2$ (fast monotone), $\xi_j>0.6$ (slow monotone),
  $-0.2<\xi_j<0$ (fast alternating), and $\xi_j<-0.6$ (slow
  alternating). By separately identifying these diverse modes, we can
  determine whether and how brain regions' differing controllability
  of particular time scales might relate to their differing
  controllability of various length scale modes.

Beginning with the control of only monotone dynamics, we find that support for slow monotone dynamics (eigenvalues $\xi_j>0.6$) differs across the brain with greatest support located in the prefrontal cortex and temporoparietal junction (Fig. \ref{fig:poscontrol}a; see a list of the highest and lowest such regions in Table \ref{tab:monfast} of the Appendix). Moreover, we find that the degree to which a regional controller supports slow monotone dynamics is strongly and positively correlated with regional average controllability (Spearman correlation coefficient $\rho=0.99$, $df=233$, $p<1\times10^{-4}$; Fig.~\ref{fig:poscontrol}c), providing support for energetically easy state transitions. Similarly, we also find that support for fast monotone dynamics (eigenvalues $0<\xi_j<0.2$) differs across the brain with greatest support located in subcortical areas (Fig. \ref{fig:poscontrol}b; see a list of the highest and lowest such regions in Table \ref{tab:monslow} of the Appendix). In addition, we find that the degree to which a regional controller supports fast monotone dynamics is strongly and positively correlated with regional modal controllability ($\rho=0.59$, $df=233$, $p<1\times10^{-4}$; Fig.~\ref{fig:poscontrol}d), providing support for energetically more difficult state transitions.

When studying the control of alternating dynamics, we find that support for slow alternating dynamics (eigenvalues $\xi_j<-0.6$) differs across the brain with greatest support located in the temporoparietal junction, inferior frontal gyrus, and precuneus (Fig. \ref{fig:negcontrol}a; see a list of the highest and lowest such regions in Table \ref{tab:altfast} of the Appendix). Moreover, we find that the degree to which a regional controller supports slow alternating dynamics is strongly and positively correlated with regional average controllability ($\rho=0.83$, $df=233$, $p<1\times10^{-4}$; Fig.~\ref{fig:negcontrol}c), providing support for energetically easy state transitions. Lastly, we also find that support for fast alternating dynamics (eigenvalues $-0.2 < \xi_j<0$) differs across the brain with greatest support located in several midline structures (Fig. \ref{fig:negcontrol}b; see a list of the highest and lowest such regions in Table \ref{tab:altslow} of the Appendix). In this case, we find that the degree to which a regional controller supports fast alternating dynamics is strongly and positively correlated with regional modal controllability ($\rho=0.24$, $df=233$, $p<2\times10^{-4}$; Fig.~\ref{fig:negcontrol}d), providing support for energetically more difficult state transitions. 

\begin{figure}
\includegraphics[width=0.95\linewidth]{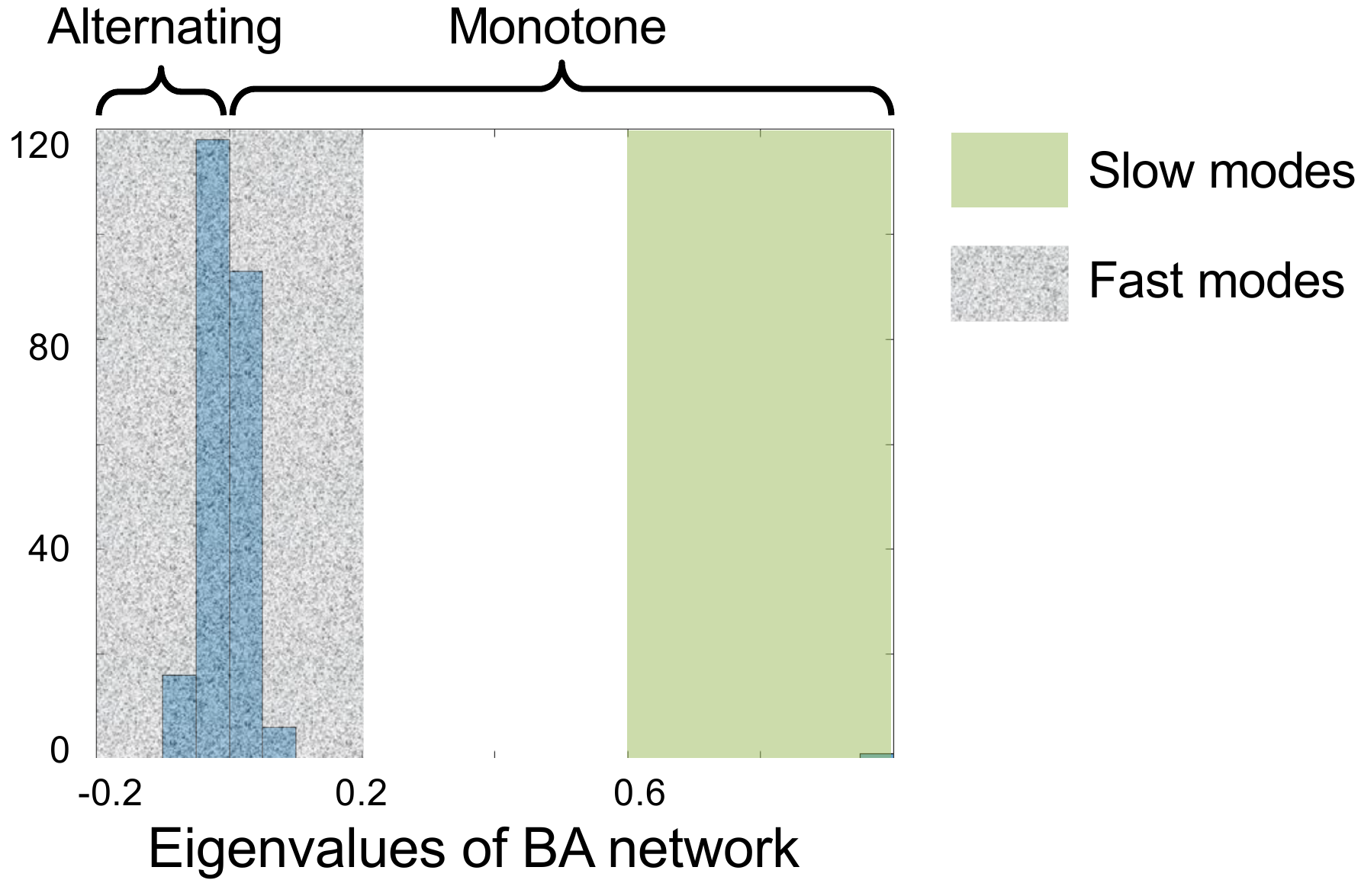}
\caption{\textbf{Histogram of $N=234$ eigenvalues of a Barabasi-Albert preferential attachment network.} As we had done for the group-representative brain network, we make partitions of $|\xi_j|<0.2$, $0.2<|\xi_j|<0.6$, and $|\xi_j|>0.6$ for positive and negative eigenvalues, respectively, which correspond to the monotone and alternating modes of the system.} \label{fig:baspectrum}
\end{figure}

\begin{figure*}
\includegraphics[width=0.7\linewidth]{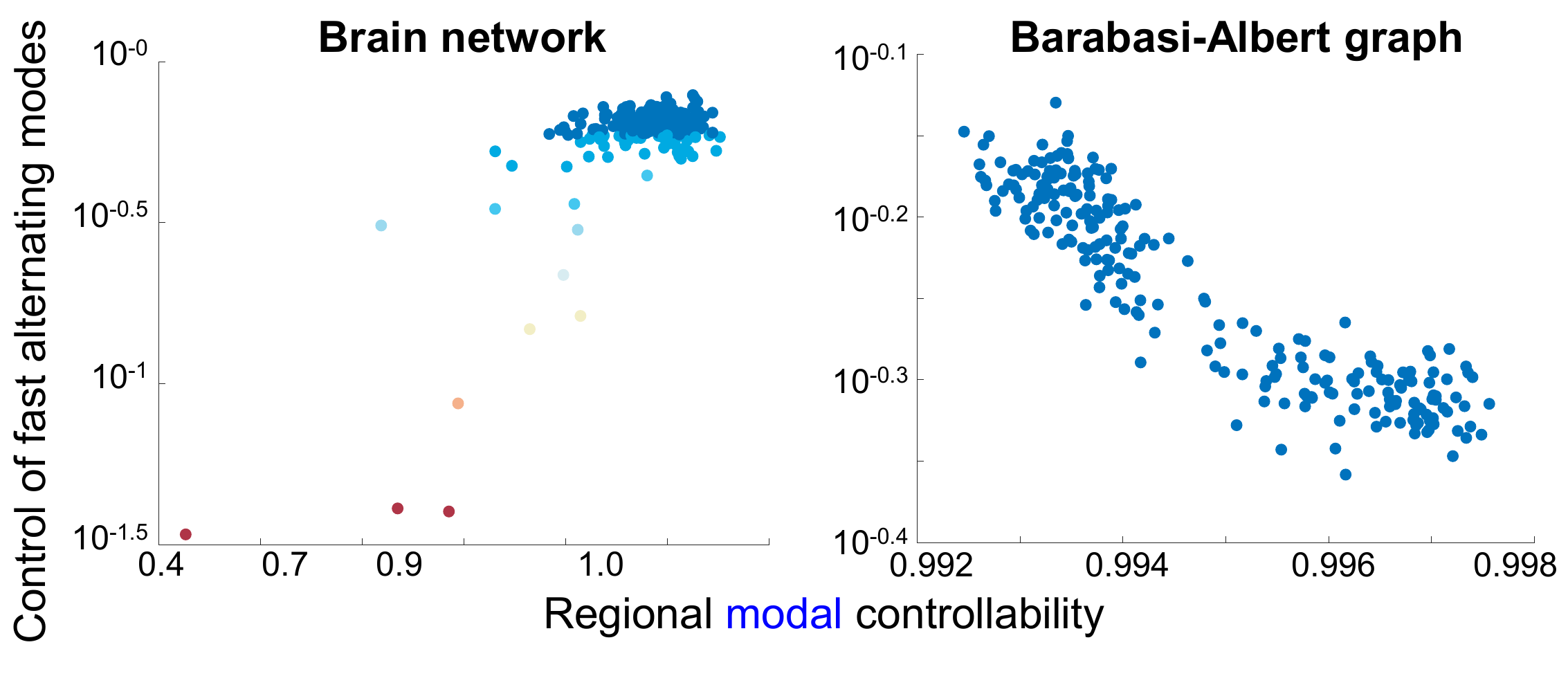}
\caption{\textbf{Different results from the group-representative brain network and a Barabasi-Albert network.} Study of the alternating dynamical modes for the two different networks shows that while controllers of fast modes are positively correlated with modal controllers in the brain network (\emph{left}), the same controllers are negatively correlated in a Barabasi-Albert network (\emph{right}), $\rho=-0.91$, $df=233$, $p<1\times10^{-4}$. (The color in the left image simply corresponds to the strength of control of fast alternating modes, for consistency with Fig. 5d.)} \label{fig:twonetworks}
\end{figure*}

\section{Dependence Upon Graph Architecture}
\label{sec:topology}

The results described thus far provide striking relations between the temporal and spatial scales of network dynamics that a brain region can control. An interesting and important question to ask is whether and to what degree these relations hold true in graph models with vastly different topologies. In choosing a particularly dissimilar comparator, we consider the fact that human brain networks are known to display high clustering and short path length \cite{bassett2016small,bassett2006small}, community structure \cite{baum2017modular}, and core-periphery structure \cite{betzel2018diversity}. A notable graph model that stands in stark contrast to this architecture is the scale free model \cite{albert2002statistical}, also sometimes referred to as the preferential attachment model \cite{price1976general} or the Barabasi-Albert model \cite{barabasi1999emergence}, which displays low clustering and longer path length, less community structure, and less core-periphery structure. For ease of comparison, we provide a table contrasting these primary network metrics in both networks, in the Appendix (see Table VII).

We therefore study a synthetically constructed Barabasi-Albert network with the same number of nodes ($N=234$) and the same average strength as the group-representative brain network in our empirical data set. As before, we calculate controllability metrics and the Laplacian eigenvectors to investigate the spatial harmonics. We find that the relationships between controllers of different length scales with average and modal controllers remain similar across both brain and Barabasi-Albert graphs. Specifically, the nodes in the Barabasi-Albert graph most relevant for the large-scale mode $|\phi_{1}^j|$ tend to be the nodes with high modal controllability: $\rho=0.64$, $df=233$, $p<1\times10^{-4}$. Meanwhile, the nodes in the Barabasi-Albert graph most relevant for the small-scale mode $|\phi_{N-1}^j|$ tend to be the nodes with high average controllability: $\rho=0.77$, $df=233$, $p<1\times10^{-5}$. Such positive correlations were also found in the brain network (Section III).

To study the control of modes with different time-scales, we employ the same partition that we used in the brain network, i.e. groups of $|\xi_j|<0.2$, $0.2<|\xi_j|<0.6$, and $|\xi_j|>0.6$ for both positive and negative eigenvalues, respectively (see Fig. \ref{fig:baspectrum}). Interestingly, we find that the distribution of eigenvalues differs significantly in the Barabasi-Albert network in comparison to the group-representative brain network, with the striking absence of slow alternating modes. When we consider only the fast alternating modes, we observe that the relationship between the controllers of these fast modes and the modal controllers is inverted in the Barabasi-Albert network in comparison to the group-representative brain network (see Fig. \ref{fig:twonetworks}). In the brain, the degree to which a regional controller supports fast alternating dynamics tracks the degree to which that same regional controller supports energetically more difficult transitions. In the Barabasi-Albert network, the degree to which a regional controller supports fast alternating dynamics is anti-correlated with the degree to which that same regional controller supports energetically more difficult transitions (Spearman correlation coefficient $\rho=-0.91$, $df=233$, $p<1\times10^{-4}$).

\section{Discussion}
\label{sec:discussion}

The human brain displays a variety of different types of dynamics. Here we use two different models to compare the control profiles of various brain regions: diffusion dynamics and network control, respectively. The two dynamical models we use capture different properties of the same brain network, hence it is useful to compare and contrast these different dynamics produced by the same system. The model for synchronizability uses the Master Stability Function (MSF) formalism based on a model of connected oscillators, whereas our controllability analysis relies on a linear discrete time model. Note that while the MSF approach originally studied continuous time systems, there have been extensions to discrete time systems \cite{Porfiri_2011}. In the latter case, a modified graph Laplacian is used that has the same eigenvalues as a regular Laplacian operator (and the eigenvectors are similar up to a constant shift) \cite{Sun_2009}. Hence, our two models are related by the structure of the network, which appears in the interconnections among the oscillators in the first model, and as a weighted adjacency matrix in the second model. Through these analyses we find that brain regions with high average controllability are also predicted to effectively control dynamics on short length scales, while regions of high modal controllability are predicted to do the converse. In addition, regions of high average controllability are also predicted to effectively control dynamics on slow scales, while regions of high modal controllability are predicted to do the converse. Collectively, our results expand our understanding of the relations between brain network dynamics and control, and the relative roles that different brain regions can play in controlling the diverse length and time scales of neural dynamics.

\textbf{Pertinent Theoretical Considerations.} There are two important theoretical considerations relevant to our work: the relation between linear and nonlinear network control, and the nature of the Master Stability Function. \rev{The first consideration is crucial, particularly given recent results regarding the difficulty of applying linear control models to nonlinear systems \cite{Jiang2019}. While linear models are certainly valid only within the domain of linearization, we note that the} modelling of brain activity in large-scale regional networks shows that the linear approximation provides fair explanatory power of resting state fMRI BOLD data \cite{Honey2009}. Further, studies of this controllability framework using non-linear oscillators connected with coupling constants estimated from large-scale white matter structural connections shows a good overlap with the linear approximation \cite{muldoon2016stimulation}. While the model we employ is a discrete-time system, this controllability Gramian is statistically similar to that obtained in a continuous-time system \cite{gu2015controllability}, through the comparison of simulations run using MATLAB's \textit{lyap} function. Regarding the second, we note that we have plotted a typical example of a Master Stability Function (MSF) for a network of oscillators schematically in Fig. \ref{fig:schematic}d; however, specific details will depend on the dynamics of individual nodes and the connectivity between them. The shape of the MSF for various families of dynamical systems is typically convex for generic oscillator systems, including chaotic oscillators that have stable limit cycles \cite{PhysRevE.80.036204}.

Separately, it is worth noting that there are several important and interesting questions at the intersection of control theory and network neuroscience. First, early work focused on the question of whether a brain network is controllable from a single region, providing initial evidence in the affirmative \cite{gu2015controllability}. A subsequent numerical study cast doubt on that evidence \cite{tu2018warnings}, while two even later studies provided additional theoretical evidence in the affirmative \cite{pasqualetti2019warnings,menara2017structural}. Notably, the structural controllability of networks depends upon their topology, being quite different for diverse graph models including the Erd\H{o}s-R\'{e}nyi, ring lattice, Watts-Strogatz, modular, random geometric, and Barabasi-Albert model \cite{pasqualetti2019warnings,wuyan2018benchmarking}, the latter of which we also study here. Second, several studies address the question of whether a set of control points (network nodes) can be identified that can effectively and reliably induce a particular change in system dynamics \cite{yan2017network,zanudo2017structure}. Third, another body of work focuses on the question of how to guide the brain from one state of activity to another state of activity given the underlying structural connections along which activity can propagate \cite{gu2017optimal,betzel2016optimally}. Work in this vein has offered validated predictions of the effects of brain stimulation \cite{stiso2019white}, and explanations of individual differences in cognitive control \cite{cui2019individual}. This list of questions is far from exhaustive, and we expand upon it in this work by probing the time scales and spatial scales relevant for control in brain networks.

\textbf{Future Directions.} Our work provides an initial foray into the combined study of brain network dynamics and control, and how the two depend upon the length scales and time scales of brain states and state transitions. From a mathematical perspective, it would be interesting to study these relations in other graph models with different topologies, and also to work to find analytical forms for the relations we observe in numerical experiments. From a neurophysics perspective, future efforts could extend our work into other age groups, or into clinical groups of patients with psychiatric disease or neurological disorders where such relations might be altered. Efforts could also determine whether our findings in humans are recapitulated in other animals, and across different spatial resolutions of imaging data. Another avenue for further investigation could be the use of directed brain networks, such as those that are available from tract-tracing studies of macaques, which would display complex eigenvalues thereby providing additional oscillatory time-scales for more extensive study. Lastly, this trimodal nature of the data (see Fig. 3) appears to arise when studying group-representative data, which led to our choice of time-scale bins. Future work could extend our study to the single-subject level, where a distinct and subject-specific binning procedure might prove helpful.

\section{Acknowledgements}

D.S.B. and E.T. acknowledge support from the John D. and Catherine T. MacArthur Foundation, the Alfred P. Sloan Foundation, the ISI Foundation, the Paul Allen Foundation, the Army Research Laboratory (W911NF-10-2-0022), the Army Research Office (Bassett-W911NF-14-1-0679, Grafton-W911NF-16-1-0474, DCIST- W911NF-17-2-0181), the Office of Naval Research, the National Institute of Mental Health (2-R01-DC-009209-11, R01 – MH112847, R01-MH107235, R21-M MH-106799), the National Institute of Child Health and Human Development (1R01HD086888-01), National Institute of Neurological Disorders and Stroke (R01 NS099348), and the National Science Foundation (BCS-1441502, BCS-1430087, NSF PHY-1554488 and BCS-1631550). The content is solely the responsibility of the authors and does not necessarily represent the official views of any of the funding agencies.

\clearpage
\newpage
\section{Appendix}
\label{sec:appendix}

All data were acquired from the Philadelphia Neurodevelopmental Cohort (PNC), a large community-based study of brain development. This resource is publicly available through the Database of Genotypes and Phenotypes. Each subject provided their informed consent according to the Institutional Review Board of the University of Pennsylvania who approved all study protocols. These subjects had no gross radiological abnormalities that distorted brain anatomy, no history of inpatient psychiatric hospitalization, no use of psychotropic medications at the time of scanning, and no medical disorders that could impact brain function. Each included subject also passed both manual and automated quality-assessment protocols for DTI \cite{Roalf2016903} and T1-weighted structural imaging \cite{Vandekar14012015}, and had low in-scanner head motion (less than 2mm mean relative displacement between \textit{b}=0 volumes). Here we study a subset of the full group previously reported in \cite{tang2017developmental}; specifically, we consider the 190 young adults aged 18 to 22 years.

Structural connectivity was estimated using 64-direction DTI data. The diffusion tensor was estimated and deterministic whole-brain fiber tracking was implemented in DSI Studio using a modified FACT algorithm, with exactly 1,000,000 streamlines initiated per subject after removing all streamlines with length less than 10mm \cite{gu2015controllability}. A 234-region parcellation \cite{10.1371/journal.pone.0048121} was constructed from the T1 image using FreeSurfer. Parcels were dilated by 4mm to extend regions into white matter, and registered to the first non-weighted (\textit{b}=0) volume using an affine transform. Edge weights $A_{ij}$ in the adjacency matrix were defined by the number of streamlines connecting each pair of nodes end-to-end. All analyses were replicated using an alternative edge weight definition, where weights are equal to the number of streamlines connecting each node pair divided by the total volume of the node pair, as well as using probabilistic fiber tracking methods. A schematic for structural connectome construction is depicted in Fig. 1b.

\begin{table}[h]
 \centering
\caption {\textbf{Brain regions with highest and lowest weights for the largest-scale Laplacian eigenvector (Fig. \ref{fig:spatial}a})}.
\begin{tabular}{|c|c|}
%\hline
%\textbf{Brain regions } & & \\
%\hline
\hline
Highest & Lowest  \\
\hline
Postcentral 6 (L) & Brainstem (L)  \\
Supramarginal 2 (L) & Precentral 1 (L)  \\
Supramarginal 3 (L) & Superior frontal 8 (L)  \\
Supramarginal 1 (L) & Caudate (L)  \\
Postcentral 7 (L) & Postcentral 1 (L)  \\
\hline
\end{tabular} \label{tab:lap1}
\end{table}

\begin{table}[h]
 \centering
\caption {\textbf{Brain regions with highest and lowest weights for the smallest-scale Laplacian eigenvector (Fig. \ref{fig:spatial}b})}.
\begin{tabular}{|c|c|}
\hline
Highest & Lowest  \\
\hline
Brainstem (L) & Postcentral 6 (L)  \\
Precentral 6 (R) & Postcentral 7 (L)  \\
Postcentral 5 (R) & Supramarginal 3 (L) \\
Thalamus proper (R) & Supramarginal 2 (L)  \\
Thalamus proper (L) & Postcentral 1 (R)  \\
\hline
\end{tabular} \label{tab:lap234}
\end{table}

\begin{table}[h]
 \centering
\caption {\textbf{Brain regions that provide the highest and lowest support for the control of slow monotone dynamics (Fig. \ref{fig:poscontrol}a})}.
\begin{tabular}{|c|c|}
\hline
Highest & Lowest  \\
\hline
Brainstem (L) & Postcentral 6 (L)  \\
Precentral 6 (R) & Superior temporal 1 (L)  \\
Thalamus proper (R) & Supramarginal 3 (L) \\
Thalamus proper (L) & Banks STS 1 (L)  \\
Superior frontal 8 (L) & Transverse temporal 1 (R)  \\
\hline
\end{tabular} \label{tab:monfast}
\end{table}

\begin{table}[h]
 \centering
\caption {\textbf{Brain regions that provide the highest and lowest support for the control of fast monotone dynamics (Fig. \ref{fig:poscontrol}b})}.
\begin{tabular}{|c|c|}
\hline
Highest & Lowest  \\
\hline
Superior frontal 8 (R) & Brainstem (L)  \\
Middle temporal 4 (L) & Superior frontal 8 (L)  \\
Insula 2 (L) & Thalamus proper (L) \\
Transverse temporal 1 (L) & Thalamus proper (R) \\
Superior temporal 2 (R) & Precentral 6 (R)  \\
\hline
\end{tabular} \label{tab:monslow}
\end{table}

\begin{table}[h]
 \centering
\caption {\textbf{Brain regions that provide the highest and lowest support for the control of slow alternating dynamics (Fig. \ref{fig:negcontrol}a})}.
\begin{tabular}{|c|c|}
\hline
Highest & Lowest  \\
\hline
Brainstem (L) & Rostral middle frontal 2 (L)  \\
Precentral 6 (R) & Lateral orbitofrontal 1 (L)  \\
Postcentral 5 (R) & Precuneus 2 (R) \\
Precentral 3 (L) & Banks STS 1 (L)  \\
Precentral 2 (L) & Inferior parietal 3 (L)  \\
\hline
\end{tabular} \label{tab:altfast}
\end{table}

\begin{table}[h]
 \centering
\caption {\textbf{Brain regions that provide the highest and lowest support for the control of fast alternating dynamics (Fig. \ref{fig:negcontrol}b})}.
\begin{tabular}{|c|c|}
\hline
Highest & Lowest  \\
\hline
Lateral occipital 3 (R) & Brainstem (L)  \\
Lateral occipital 1 (R) & Thalamus proper (L)  \\
Precuneus 2 (L) & Superior frontal 8 (L) \\
Lingual 3 (L) & Thalamus proper (R)  \\
Accumbens area (L) & Superior frontal 5 (R)  \\
\hline
\end{tabular} \label{tab:altslow}
\end{table}

\begin{table}[h]
\begin{tabular}{|l|l|l|}
\hline
& Brain network  & Barabasi-Albert graph \\ \hline
Clustering coefficient \cite{watts1998} & $(5.5\x10^{-4}) \pm (1.5\x10^{-4})$ & $(4.3\x10^{-3}) \pm (1.3\x10^{-5})$ \\ \hline
Characteristic path length & $0.75$ & $0.75$ \\
(binary) \cite{watts1998} & & \\ \hline
Characteristic path length & $2.1\x10^{-3}$ & $4.1\x10^{-3}$ \\
(weighted) \cite{rubinov2010} & & \\ \hline
Modularity \cite{newman2004} & $0.49 \pm (1.6\x10^{-4})$ & $0.021 \pm (2.5\x10^{-5})$ \\ \hline
Coreness \cite{borgatti2000} & $0.37$ & $0.22$ \\ \hline
\end{tabular}
\caption{\label{tab:network} (Table VII in the manuscript.) We calculate primary network metrics for both the brain network and a Barabasi-Albert graph with the same number of nodes ($N=234$) and the same normalization (see Section I.B): the metrics of clustering coefficient, characteristic path length, modularity, and coreness, respectively \cite{rubinov2010}. Uncertainty measurements are standard errors; modularity (a heuristic) is calculated over 1000 trials maximizing a common modularity quality function \cite{newman2004}.}
\end{table}

\newpage
\clearpage
\bibliography{scales,braincontrol,bibfile,bibfile_new,bibfile_original,bibfile_2,dynamical_trajectories_v5_discussionOnly_db,dynamical_trajectories_v5_structureFunction_RB,control_sync_DB,library}

\begin{thebibliography}{90}%
\makeatletter
\providecommand \@ifxundefined [1]{%
 \@ifx{#1\undefined}
}%
\providecommand \@ifnum [1]{%
 \ifnum #1\expandafter \@firstoftwo
 \else \expandafter \@secondoftwo
 \fi
}%
\providecommand \@ifx [1]{%
 \ifx #1\expandafter \@firstoftwo
 \else \expandafter \@secondoftwo
 \fi
}%
\providecommand \natexlab [1]{#1}%
\providecommand \enquote  [1]{``#1''}%
\providecommand \bibnamefont  [1]{#1}%
\providecommand \bibfnamefont [1]{#1}%
\providecommand \citenamefont [1]{#1}%
\providecommand \href@noop [0]{\@secondoftwo}%
\providecommand \href [0]{\begingroup \@sanitize@url \@href}%
\providecommand \@href[1]{\@@startlink{#1}\@@href}%
\providecommand \@@href[1]{\endgroup#1\@@endlink}%
\providecommand \@sanitize@url [0]{\catcode `\\12\catcode `\$12\catcode
  `\&12\catcode `\#12\catcode `\^12\catcode `\_12\catcode `\%12\relax}%
\providecommand \@@startlink[1]{}%
\providecommand \@@endlink[0]{}%
\providecommand \url  [0]{\begingroup\@sanitize@url \@url }%
\providecommand \@url [1]{\endgroup\@href {#1}{\urlprefix }}%
\providecommand \urlprefix  [0]{URL }%
\providecommand \Eprint [0]{\href }%
\providecommand \doibase [0]{http://dx.doi.org/}%
\providecommand \selectlanguage [0]{\@gobble}%
\providecommand \bibinfo  [0]{\@secondoftwo}%
\providecommand \bibfield  [0]{\@secondoftwo}%
\providecommand \translation [1]{[#1]}%
\providecommand \BibitemOpen [0]{}%
\providecommand \bibitemStop [0]{}%
\providecommand \bibitemNoStop [0]{.\EOS\space}%
\providecommand \EOS [0]{\spacefactor3000\relax}%
\providecommand \BibitemShut  [1]{\csname bibitem#1\endcsname}%
\let\auto@bib@innerbib\@empty
%</preamble>
\bibitem [{\citenamefont {Bassett}\ and\ \citenamefont
  {Sporns}(2017)}]{bassett2017network}%
  \BibitemOpen
  \bibfield  {author} {\bibinfo {author} {\bibfnamefont {D.~S.}\ \bibnamefont
  {Bassett}}\ and\ \bibinfo {author} {\bibfnamefont {O.}~\bibnamefont
  {Sporns}},\ }\href@noop {} {\bibfield  {journal} {\bibinfo  {journal} {Nat
  Neurosci}\ }\textbf {\bibinfo {volume} {20}},\ \bibinfo {pages} {353}
  (\bibinfo {year} {2017})}\BibitemShut {NoStop}%
\bibitem [{\citenamefont {Breakspear}(2017)}]{breakspear2017dynamic}%
  \BibitemOpen
  \bibfield  {author} {\bibinfo {author} {\bibfnamefont {M.}~\bibnamefont
  {Breakspear}},\ }\href@noop {} {\bibfield  {journal} {\bibinfo  {journal}
  {Nat Neurosci}\ }\textbf {\bibinfo {volume} {20}},\ \bibinfo {pages} {340}
  (\bibinfo {year} {2017})}\BibitemShut {NoStop}%
\bibitem [{\citenamefont {Bansal}\ \emph {et~al.}(2018)\citenamefont {Bansal},
  \citenamefont {Nakuci},\ and\ \citenamefont
  {Muldoon}}]{bansal2018personalized}%
  \BibitemOpen
  \bibfield  {author} {\bibinfo {author} {\bibfnamefont {K.}~\bibnamefont
  {Bansal}}, \bibinfo {author} {\bibfnamefont {J.}~\bibnamefont {Nakuci}}, \
  and\ \bibinfo {author} {\bibfnamefont {S.~F.}\ \bibnamefont {Muldoon}},\
  }\href@noop {} {\bibfield  {journal} {\bibinfo  {journal} {Curr Opin
  Neurobiol}\ }\textbf {\bibinfo {volume} {52}},\ \bibinfo {pages} {42}
  (\bibinfo {year} {2018})}\BibitemShut {NoStop}%
\bibitem [{\citenamefont {Proix}\ \emph {et~al.}(2017)\citenamefont {Proix},
  \citenamefont {Bartolomei}, \citenamefont {Guye},\ and\ \citenamefont
  {Jirsa}}]{proix2017individual}%
  \BibitemOpen
  \bibfield  {author} {\bibinfo {author} {\bibfnamefont {T.}~\bibnamefont
  {Proix}}, \bibinfo {author} {\bibfnamefont {F.}~\bibnamefont {Bartolomei}},
  \bibinfo {author} {\bibfnamefont {M.}~\bibnamefont {Guye}}, \ and\ \bibinfo
  {author} {\bibfnamefont {V.~K.}\ \bibnamefont {Jirsa}},\ }\href@noop {}
  {\bibfield  {journal} {\bibinfo  {journal} {Brain}\ }\textbf {\bibinfo
  {volume} {140}},\ \bibinfo {pages} {641} (\bibinfo {year}
  {2017})}\BibitemShut {NoStop}%
\bibitem [{\citenamefont {Falcon}\ \emph {et~al.}(2016)\citenamefont {Falcon},
  \citenamefont {Jirsa},\ and\ \citenamefont {Solodkin}}]{falcon2016new}%
  \BibitemOpen
  \bibfield  {author} {\bibinfo {author} {\bibfnamefont {M.~I.}\ \bibnamefont
  {Falcon}}, \bibinfo {author} {\bibfnamefont {V.}~\bibnamefont {Jirsa}}, \
  and\ \bibinfo {author} {\bibfnamefont {A.}~\bibnamefont {Solodkin}},\
  }\href@noop {} {\bibfield  {journal} {\bibinfo  {journal} {Curr Opin Neurol}\
  }\textbf {\bibinfo {volume} {29}},\ \bibinfo {pages} {429} (\bibinfo {year}
  {2016})}\BibitemShut {NoStop}%
\bibitem [{\citenamefont {Schirner}\ \emph {et~al.}(2015)\citenamefont
  {Schirner}, \citenamefont {Rothmeier}, \citenamefont {Jirsa}, \citenamefont
  {McIntosh},\ and\ \citenamefont {Ritter}}]{ritter2015automated}%
  \BibitemOpen
  \bibfield  {author} {\bibinfo {author} {\bibfnamefont {M.}~\bibnamefont
  {Schirner}}, \bibinfo {author} {\bibfnamefont {S.}~\bibnamefont {Rothmeier}},
  \bibinfo {author} {\bibfnamefont {V.~K.}\ \bibnamefont {Jirsa}}, \bibinfo
  {author} {\bibfnamefont {A.~R.}\ \bibnamefont {McIntosh}}, \ and\ \bibinfo
  {author} {\bibfnamefont {P.}~\bibnamefont {Ritter}},\ }\href@noop {}
  {\bibfield  {journal} {\bibinfo  {journal} {Neuroimage}\ }\textbf {\bibinfo
  {volume} {117}},\ \bibinfo {pages} {343} (\bibinfo {year}
  {2015})}\BibitemShut {NoStop}%
\bibitem [{\citenamefont {Bassett}\ \emph {et~al.}(2018)\citenamefont
  {Bassett}, \citenamefont {Zurn},\ and\ \citenamefont {Gold}}]{bassett2018on}%
  \BibitemOpen
  \bibfield  {author} {\bibinfo {author} {\bibfnamefont {D.~S.}\ \bibnamefont
  {Bassett}}, \bibinfo {author} {\bibfnamefont {P.}~\bibnamefont {Zurn}}, \
  and\ \bibinfo {author} {\bibfnamefont {J.~I.}\ \bibnamefont {Gold}},\
  }\href@noop {} {\bibfield  {journal} {\bibinfo  {journal} {Nat Rev Neurosci}\
  }\textbf {\bibinfo {volume} {Epub Ahead of Print}} (\bibinfo {year}
  {2018})}\BibitemShut {NoStop}%
\bibitem [{\citenamefont {Breakspear}\ and\ \citenamefont
  {Stam}(2005)}]{breakspear2005dynamics}%
  \BibitemOpen
  \bibfield  {author} {\bibinfo {author} {\bibfnamefont {M.}~\bibnamefont
  {Breakspear}}\ and\ \bibinfo {author} {\bibfnamefont {C.~J.}\ \bibnamefont
  {Stam}},\ }\href@noop {} {\bibfield  {journal} {\bibinfo  {journal} {Philos
  Trans R Soc Lond B Biol Sci}\ }\textbf {\bibinfo {volume} {360}},\ \bibinfo
  {pages} {1051} (\bibinfo {year} {2005})}\BibitemShut {NoStop}%
\bibitem [{\citenamefont {Betzel}\ and\ \citenamefont
  {Bassett}(2017)}]{betzel2017multi}%
  \BibitemOpen
  \bibfield  {author} {\bibinfo {author} {\bibfnamefont {R.~F.}\ \bibnamefont
  {Betzel}}\ and\ \bibinfo {author} {\bibfnamefont {D.~S.}\ \bibnamefont
  {Bassett}},\ }\href@noop {} {\bibfield  {journal} {\bibinfo  {journal}
  {Neuroimage}\ }\textbf {\bibinfo {volume} {160}},\ \bibinfo {pages} {73}
  (\bibinfo {year} {2017})}\BibitemShut {NoStop}%
\bibitem [{\citenamefont {Friston}(2005)}]{friston2005models}%
  \BibitemOpen
  \bibfield  {author} {\bibinfo {author} {\bibfnamefont {K.~J.}\ \bibnamefont
  {Friston}},\ }\href@noop {} {\bibfield  {journal} {\bibinfo  {journal} {Annu
  Rev Psychol}\ }\textbf {\bibinfo {volume} {56}},\ \bibinfo {pages} {57}
  (\bibinfo {year} {2005})}\BibitemShut {NoStop}%
\bibitem [{\citenamefont {Friston}(2008)}]{friston2008hierarchical}%
  \BibitemOpen
  \bibfield  {author} {\bibinfo {author} {\bibfnamefont {K.}~\bibnamefont
  {Friston}},\ }\href@noop {} {\bibfield  {journal} {\bibinfo  {journal} {PLoS
  Comput Biol}\ }\textbf {\bibinfo {volume} {4}},\ \bibinfo {pages} {e1000211}
  (\bibinfo {year} {2008})}\BibitemShut {NoStop}%
\bibitem [{\citenamefont {Deco}\ \emph {et~al.}(2008)\citenamefont {Deco},
  \citenamefont {Jirsa}, \citenamefont {Robinson}, \citenamefont {Breakspear},\
  and\ \citenamefont {Friston}}]{deco2008dynamic}%
  \BibitemOpen
  \bibfield  {author} {\bibinfo {author} {\bibfnamefont {G.}~\bibnamefont
  {Deco}}, \bibinfo {author} {\bibfnamefont {V.~K.}\ \bibnamefont {Jirsa}},
  \bibinfo {author} {\bibfnamefont {P.~A.}\ \bibnamefont {Robinson}}, \bibinfo
  {author} {\bibfnamefont {M.}~\bibnamefont {Breakspear}}, \ and\ \bibinfo
  {author} {\bibfnamefont {K.}~\bibnamefont {Friston}},\ }\href@noop {}
  {\bibfield  {journal} {\bibinfo  {journal} {PLoS Comput Biol}\ }\textbf
  {\bibinfo {volume} {4}},\ \bibinfo {pages} {e1000092} (\bibinfo {year}
  {2008})}\BibitemShut {NoStop}%
\bibitem [{\citenamefont {Honey}\ \emph {et~al.}(2009)\citenamefont {Honey},
  \citenamefont {Sporns}, \citenamefont {Cammoun}, \citenamefont {Gigandet},
  \citenamefont {Thiran}, \citenamefont {Meuli},\ and\ \citenamefont
  {Hagmann}}]{Honey2009}%
  \BibitemOpen
  \bibfield  {author} {\bibinfo {author} {\bibfnamefont {C.~J.}\ \bibnamefont
  {Honey}}, \bibinfo {author} {\bibfnamefont {O.}~\bibnamefont {Sporns}},
  \bibinfo {author} {\bibfnamefont {L.}~\bibnamefont {Cammoun}}, \bibinfo
  {author} {\bibfnamefont {X.}~\bibnamefont {Gigandet}}, \bibinfo {author}
  {\bibfnamefont {J.~P.}\ \bibnamefont {Thiran}}, \bibinfo {author}
  {\bibfnamefont {R.}~\bibnamefont {Meuli}}, \ and\ \bibinfo {author}
  {\bibfnamefont {P.}~\bibnamefont {Hagmann}},\ }\href@noop {} {\bibfield
  {journal} {\bibinfo  {journal} {Proc Natl Acad Sci U S A}\ }\textbf {\bibinfo
  {volume} {106}},\ \bibinfo {pages} {2035} (\bibinfo {year}
  {2009})}\BibitemShut {NoStop}%
\bibitem [{\citenamefont {Fern\'{a}ndez~Gal\'{a}n}(2008)}]{Galan2008}%
  \BibitemOpen
  \bibfield  {author} {\bibinfo {author} {\bibfnamefont {R.}~\bibnamefont
  {Fern\'{a}ndez~Gal\'{a}n}},\ }\href@noop {} {\bibfield  {journal} {\bibinfo
  {journal} {PLoS One}\ }\textbf {\bibinfo {volume} {3}},\ \bibinfo {pages}
  {e2148} (\bibinfo {year} {2008})}\BibitemShut {NoStop}%
\bibitem [{\citenamefont {Raj}\ \emph {et~al.}(2012)\citenamefont {Raj},
  \citenamefont {Kuceyeski},\ and\ \citenamefont {Weiner}}]{Raj2012}%
  \BibitemOpen
  \bibfield  {author} {\bibinfo {author} {\bibfnamefont {A.}~\bibnamefont
  {Raj}}, \bibinfo {author} {\bibfnamefont {A.}~\bibnamefont {Kuceyeski}}, \
  and\ \bibinfo {author} {\bibfnamefont {M.}~\bibnamefont {Weiner}},\ }\href
  {\doibase 10.1016/j.neuron.2011.12.040} {\bibfield  {journal} {\bibinfo
  {journal} {Neuron}\ }\textbf {\bibinfo {volume} {73}},\ \bibinfo {pages}
  {1204} (\bibinfo {year} {2012})}\BibitemShut {NoStop}%
\bibitem [{\citenamefont {Aquino}\ \emph {et~al.}(2012)\citenamefont {Aquino},
  \citenamefont {Schira}, \citenamefont {Robinson}, \citenamefont {Drysdale},\
  and\ \citenamefont {Breakspear}}]{10.1371/journal.pcbi.1002435}%
  \BibitemOpen
  \bibfield  {author} {\bibinfo {author} {\bibfnamefont {K.~M.}\ \bibnamefont
  {Aquino}}, \bibinfo {author} {\bibfnamefont {M.~M.}\ \bibnamefont {Schira}},
  \bibinfo {author} {\bibfnamefont {P.~A.}\ \bibnamefont {Robinson}}, \bibinfo
  {author} {\bibfnamefont {P.~M.}\ \bibnamefont {Drysdale}}, \ and\ \bibinfo
  {author} {\bibfnamefont {M.}~\bibnamefont {Breakspear}},\ }\href
  {https://doi.org/10.1371/journal.pcbi.1002435} {\bibfield  {journal}
  {\bibinfo  {journal} {PLOS Computational Biology}\ }\textbf {\bibinfo
  {volume} {8}},\ \bibinfo {pages} {1} (\bibinfo {year} {2012})}\BibitemShut
  {NoStop}%
\bibitem [{\citenamefont {Viventi}\ \emph {et~al.}(2011)\citenamefont
  {Viventi}, \citenamefont {Kim}, \citenamefont {Vigeland}, \citenamefont
  {Frechette}, \citenamefont {Blanco}, \citenamefont {Kim}, \citenamefont
  {Avrin}, \citenamefont {Tiruvadi}, \citenamefont {Hwang}, \citenamefont
  {Vanleer}, \citenamefont {Wulsin}, \citenamefont {Davis}, \citenamefont
  {Gelber}, \citenamefont {Palmer}, \citenamefont {Van~der Spiegel},
  \citenamefont {Wu}, \citenamefont {Xiao}, \citenamefont {Huang},
  \citenamefont {Contreras}, \citenamefont {Rogers},\ and\ \citenamefont
  {Litt}}]{Viventi2011}%
  \BibitemOpen
  \bibfield  {author} {\bibinfo {author} {\bibfnamefont {J.}~\bibnamefont
  {Viventi}}, \bibinfo {author} {\bibfnamefont {D.-H.}\ \bibnamefont {Kim}},
  \bibinfo {author} {\bibfnamefont {L.}~\bibnamefont {Vigeland}}, \bibinfo
  {author} {\bibfnamefont {E.~S.}\ \bibnamefont {Frechette}}, \bibinfo {author}
  {\bibfnamefont {J.~A.}\ \bibnamefont {Blanco}}, \bibinfo {author}
  {\bibfnamefont {Y.-S.}\ \bibnamefont {Kim}}, \bibinfo {author} {\bibfnamefont
  {A.~E.}\ \bibnamefont {Avrin}}, \bibinfo {author} {\bibfnamefont {V.~R.}\
  \bibnamefont {Tiruvadi}}, \bibinfo {author} {\bibfnamefont {S.-W.}\
  \bibnamefont {Hwang}}, \bibinfo {author} {\bibfnamefont {A.~C.}\ \bibnamefont
  {Vanleer}}, \bibinfo {author} {\bibfnamefont {D.~F.}\ \bibnamefont {Wulsin}},
  \bibinfo {author} {\bibfnamefont {K.}~\bibnamefont {Davis}}, \bibinfo
  {author} {\bibfnamefont {C.~E.}\ \bibnamefont {Gelber}}, \bibinfo {author}
  {\bibfnamefont {L.}~\bibnamefont {Palmer}}, \bibinfo {author} {\bibfnamefont
  {J.}~\bibnamefont {Van~der Spiegel}}, \bibinfo {author} {\bibfnamefont
  {J.}~\bibnamefont {Wu}}, \bibinfo {author} {\bibfnamefont {J.}~\bibnamefont
  {Xiao}}, \bibinfo {author} {\bibfnamefont {Y.}~\bibnamefont {Huang}},
  \bibinfo {author} {\bibfnamefont {D.}~\bibnamefont {Contreras}}, \bibinfo
  {author} {\bibfnamefont {J.~A.}\ \bibnamefont {Rogers}}, \ and\ \bibinfo
  {author} {\bibfnamefont {B.}~\bibnamefont {Litt}},\ }\href
  {http://dx.doi.org/10.1038/nn.2973} {\bibfield  {journal} {\bibinfo
  {journal} {Nature Neuroscience}\ }\textbf {\bibinfo {volume} {14}},\ \bibinfo
  {pages} {1599} (\bibinfo {year} {2011})}\BibitemShut {NoStop}%
\bibitem [{\citenamefont {Kopell}\ \emph {et~al.}(2014)\citenamefont {Kopell},
  \citenamefont {Gritton}, \citenamefont {Whittington},\ and\ \citenamefont
  {Kramer}}]{kopell2014beyond}%
  \BibitemOpen
  \bibfield  {author} {\bibinfo {author} {\bibfnamefont {N.~J.}\ \bibnamefont
  {Kopell}}, \bibinfo {author} {\bibfnamefont {H.~J.}\ \bibnamefont {Gritton}},
  \bibinfo {author} {\bibfnamefont {M.~A.}\ \bibnamefont {Whittington}}, \ and\
  \bibinfo {author} {\bibfnamefont {M.~A.}\ \bibnamefont {Kramer}},\
  }\href@noop {} {\bibfield  {journal} {\bibinfo  {journal} {Neuron}\ }\textbf
  {\bibinfo {volume} {83}},\ \bibinfo {pages} {1319} (\bibinfo {year}
  {2014})}\BibitemShut {NoStop}%
\bibitem [{\citenamefont {Fries}(2015)}]{fries2015rhythms}%
  \BibitemOpen
  \bibfield  {author} {\bibinfo {author} {\bibfnamefont {P.}~\bibnamefont
  {Fries}},\ }\href@noop {} {\bibfield  {journal} {\bibinfo  {journal}
  {Neuron}\ }\textbf {\bibinfo {volume} {88}},\ \bibinfo {pages} {220}
  (\bibinfo {year} {2015})}\BibitemShut {NoStop}%
\bibitem [{\citenamefont {Palmigiano}\ \emph {et~al.}(2017)\citenamefont
  {Palmigiano}, \citenamefont {Geisel}, \citenamefont {Wolf},\ and\
  \citenamefont {Battaglia}}]{palmigiano2017flexible}%
  \BibitemOpen
  \bibfield  {author} {\bibinfo {author} {\bibfnamefont {A.}~\bibnamefont
  {Palmigiano}}, \bibinfo {author} {\bibfnamefont {T.}~\bibnamefont {Geisel}},
  \bibinfo {author} {\bibfnamefont {F.}~\bibnamefont {Wolf}}, \ and\ \bibinfo
  {author} {\bibfnamefont {D.}~\bibnamefont {Battaglia}},\ }\href@noop {}
  {\bibfield  {journal} {\bibinfo  {journal} {Nat Neurosci}\ }\textbf {\bibinfo
  {volume} {20}},\ \bibinfo {pages} {1014} (\bibinfo {year}
  {2017})}\BibitemShut {NoStop}%
\bibitem [{\citenamefont {Kirst}\ \emph {et~al.}(2016)\citenamefont {Kirst},
  \citenamefont {Timme},\ and\ \citenamefont {Battaglia}}]{kirst2016dynamic}%
  \BibitemOpen
  \bibfield  {author} {\bibinfo {author} {\bibfnamefont {C.}~\bibnamefont
  {Kirst}}, \bibinfo {author} {\bibfnamefont {M.}~\bibnamefont {Timme}}, \ and\
  \bibinfo {author} {\bibfnamefont {D.}~\bibnamefont {Battaglia}},\ }\href@noop
  {} {\bibfield  {journal} {\bibinfo  {journal} {Nat Commun}\ }\textbf
  {\bibinfo {volume} {7}},\ \bibinfo {pages} {11061} (\bibinfo {year}
  {2016})}\BibitemShut {NoStop}%
\bibitem [{\citenamefont {Muller}\ \emph {et~al.}(2018)\citenamefont {Muller},
  \citenamefont {Chavane}, \citenamefont {Reynolds},\ and\ \citenamefont
  {Sejnowski}}]{Muller2018}%
  \BibitemOpen
  \bibfield  {author} {\bibinfo {author} {\bibfnamefont {L.}~\bibnamefont
  {Muller}}, \bibinfo {author} {\bibfnamefont {F.}~\bibnamefont {Chavane}},
  \bibinfo {author} {\bibfnamefont {J.}~\bibnamefont {Reynolds}}, \ and\
  \bibinfo {author} {\bibfnamefont {T.~J.}\ \bibnamefont {Sejnowski}},\ }\href
  {http://dx.doi.org/10.1038/nrn.2018.20} {\bibfield  {journal} {\bibinfo
  {journal} {Nature Reviews Neuroscience}\ }\textbf {\bibinfo {volume} {19}},\
  \bibinfo {pages} {255} (\bibinfo {year} {2018})},\ \bibinfo {note} {review
  Article}\BibitemShut {NoStop}%
\bibitem [{\citenamefont {Heitmann}\ \emph {et~al.}(2013)\citenamefont
  {Heitmann}, \citenamefont {Boonstra},\ and\ \citenamefont
  {Breakspear}}]{10.1371/journal.pcbi.1003260}%
  \BibitemOpen
  \bibfield  {author} {\bibinfo {author} {\bibfnamefont {S.}~\bibnamefont
  {Heitmann}}, \bibinfo {author} {\bibfnamefont {T.}~\bibnamefont {Boonstra}},
  \ and\ \bibinfo {author} {\bibfnamefont {M.}~\bibnamefont {Breakspear}},\
  }\href {\doibase 10.1371/journal.pcbi.1003260} {\bibfield  {journal}
  {\bibinfo  {journal} {PLOS Computational Biology}\ }\textbf {\bibinfo
  {volume} {9}},\ \bibinfo {pages} {1} (\bibinfo {year} {2013})}\BibitemShut
  {NoStop}%
\bibitem [{\citenamefont {Huang}\ \emph {et~al.}(2010)\citenamefont {Huang},
  \citenamefont {Xu}, \citenamefont {Liang}, \citenamefont {Takagaki},
  \citenamefont {Gao},\ and\ \citenamefont {young Wu}}]{HUANG2010978}%
  \BibitemOpen
  \bibfield  {author} {\bibinfo {author} {\bibfnamefont {X.}~\bibnamefont
  {Huang}}, \bibinfo {author} {\bibfnamefont {W.}~\bibnamefont {Xu}}, \bibinfo
  {author} {\bibfnamefont {J.}~\bibnamefont {Liang}}, \bibinfo {author}
  {\bibfnamefont {K.}~\bibnamefont {Takagaki}}, \bibinfo {author}
  {\bibfnamefont {X.}~\bibnamefont {Gao}}, \ and\ \bibinfo {author}
  {\bibfnamefont {J.}~\bibnamefont {young Wu}},\ }\href {\doibase
  https://doi.org/10.1016/j.neuron.2010.11.007} {\bibfield  {journal} {\bibinfo
   {journal} {Neuron}\ }\textbf {\bibinfo {volume} {68}},\ \bibinfo {pages}
  {978 } (\bibinfo {year} {2010})}\BibitemShut {NoStop}%
\bibitem [{\citenamefont {Stiso}\ and\ \citenamefont
  {Bassett}(2018)}]{stiso2018spatial}%
  \BibitemOpen
  \bibfield  {author} {\bibinfo {author} {\bibfnamefont {J.}~\bibnamefont
  {Stiso}}\ and\ \bibinfo {author} {\bibfnamefont {D.~S.}\ \bibnamefont
  {Bassett}},\ }\href@noop {} {\bibfield  {journal} {\bibinfo  {journal}
  {arXiv}\ }\textbf {\bibinfo {volume} {1807}},\ \bibinfo {pages} {04691}
  (\bibinfo {year} {2018})}\BibitemShut {NoStop}%
\bibitem [{\citenamefont {Chaudhuri}\ \emph {et~al.}(2015)\citenamefont
  {Chaudhuri}, \citenamefont {Knoblauch}, \citenamefont {Gariel}, \citenamefont
  {Kennedy},\ and\ \citenamefont {Wang}}]{CHAUDHURI2015419}%
  \BibitemOpen
  \bibfield  {author} {\bibinfo {author} {\bibfnamefont {R.}~\bibnamefont
  {Chaudhuri}}, \bibinfo {author} {\bibfnamefont {K.}~\bibnamefont
  {Knoblauch}}, \bibinfo {author} {\bibfnamefont {M.-A.}\ \bibnamefont
  {Gariel}}, \bibinfo {author} {\bibfnamefont {H.}~\bibnamefont {Kennedy}}, \
  and\ \bibinfo {author} {\bibfnamefont {X.-J.}\ \bibnamefont {Wang}},\
  }\href@noop {} {\bibfield  {journal} {\bibinfo  {journal} {Neuron}\ }\textbf
  {\bibinfo {volume} {88}},\ \bibinfo {pages} {419 } (\bibinfo {year}
  {2015})}\BibitemShut {NoStop}%
\bibitem [{\citenamefont {Cancelli}\ \emph {et~al.}(2016)\citenamefont
  {Cancelli}, \citenamefont {Cottone}, \citenamefont {Tecchio}, \citenamefont
  {Truong}, \citenamefont {Dmochowski},\ and\ \citenamefont
  {Bikson}}]{1741-2552-13-3-036022}%
  \BibitemOpen
  \bibfield  {author} {\bibinfo {author} {\bibfnamefont {A.}~\bibnamefont
  {Cancelli}}, \bibinfo {author} {\bibfnamefont {C.}~\bibnamefont {Cottone}},
  \bibinfo {author} {\bibfnamefont {F.}~\bibnamefont {Tecchio}}, \bibinfo
  {author} {\bibfnamefont {D.~Q.}\ \bibnamefont {Truong}}, \bibinfo {author}
  {\bibfnamefont {J.}~\bibnamefont {Dmochowski}}, \ and\ \bibinfo {author}
  {\bibfnamefont {M.}~\bibnamefont {Bikson}},\ }\href
  {http://stacks.iop.org/1741-2552/13/i=3/a=036022} {\bibfield  {journal}
  {\bibinfo  {journal} {Journal of Neural Engineering}\ }\textbf {\bibinfo
  {volume} {13}},\ \bibinfo {pages} {036022} (\bibinfo {year}
  {2016})}\BibitemShut {NoStop}%
\bibitem [{\citenamefont {Lopes}\ \emph {et~al.}(2012)\citenamefont {Lopes},
  \citenamefont {Davies}, \citenamefont {Toumazou},\ and\ \citenamefont
  {Grossman}}]{6346881}%
  \BibitemOpen
  \bibfield  {author} {\bibinfo {author} {\bibfnamefont {S.}~\bibnamefont
  {Lopes}}, \bibinfo {author} {\bibfnamefont {N.}~\bibnamefont {Davies}},
  \bibinfo {author} {\bibfnamefont {C.}~\bibnamefont {Toumazou}}, \ and\
  \bibinfo {author} {\bibfnamefont {N.}~\bibnamefont {Grossman}},\ }in\ \href
  {\doibase 10.1109/EMBC.2012.6346881} {\emph {\bibinfo {booktitle} {2012
  Annual International Conference of the IEEE Engineering in Medicine and
  Biology Society}}}\ (\bibinfo {year} {2012})\ pp.\ \bibinfo {pages}
  {4152--4155}\BibitemShut {NoStop}%
\bibitem [{\citenamefont {Metwally}\ \emph {et~al.}(2012)\citenamefont
  {Metwally}, \citenamefont {Cho}, \citenamefont {Park},\ and\ \citenamefont
  {Kim}}]{6347243}%
  \BibitemOpen
  \bibfield  {author} {\bibinfo {author} {\bibfnamefont {M.~K.}\ \bibnamefont
  {Metwally}}, \bibinfo {author} {\bibfnamefont {Y.~S.}\ \bibnamefont {Cho}},
  \bibinfo {author} {\bibfnamefont {H.~J.}\ \bibnamefont {Park}}, \ and\
  \bibinfo {author} {\bibfnamefont {T.~S.}\ \bibnamefont {Kim}},\ }in\ \href
  {\doibase 10.1109/EMBC.2012.6347243} {\emph {\bibinfo {booktitle} {2012
  Annual International Conference of the IEEE Engineering in Medicine and
  Biology Society}}}\ (\bibinfo {year} {2012})\ pp.\ \bibinfo {pages}
  {5514--5517}\BibitemShut {NoStop}%
\bibitem [{\citenamefont {Raj}\ \emph {et~al.}(2015)\citenamefont {Raj},
  \citenamefont {LoCastro}, \citenamefont {Kuceyeski}, \citenamefont {Tosun},
  \citenamefont {Relkin},\ and\ \citenamefont {Weiner}}]{Raj2015}%
  \BibitemOpen
  \bibfield  {author} {\bibinfo {author} {\bibfnamefont {A.}~\bibnamefont
  {Raj}}, \bibinfo {author} {\bibfnamefont {E.}~\bibnamefont {LoCastro}},
  \bibinfo {author} {\bibfnamefont {A.}~\bibnamefont {Kuceyeski}}, \bibinfo
  {author} {\bibfnamefont {D.}~\bibnamefont {Tosun}}, \bibinfo {author}
  {\bibfnamefont {N.}~\bibnamefont {Relkin}}, \ and\ \bibinfo {author}
  {\bibfnamefont {M.}~\bibnamefont {Weiner}},\ }\href {\doibase
  10.1016/j.celrep.2014.12.034} {\bibfield  {journal} {\bibinfo  {journal}
  {Cell Reports}\ }\textbf {\bibinfo {volume} {10}},\ \bibinfo {pages} {359}
  (\bibinfo {year} {2015})}\BibitemShut {NoStop}%
\bibitem [{\citenamefont {Jones}(2016)}]{jones2016when}%
  \BibitemOpen
  \bibfield  {author} {\bibinfo {author} {\bibfnamefont {S.~R.}\ \bibnamefont
  {Jones}},\ }\href@noop {} {\bibfield  {journal} {\bibinfo  {journal} {Curr
  Opin Neurobiol}\ }\textbf {\bibinfo {volume} {40}},\ \bibinfo {pages} {72}
  (\bibinfo {year} {2016})}\BibitemShut {NoStop}%
\bibitem [{\citenamefont {Sacchet}\ \emph {et~al.}(2015)\citenamefont
  {Sacchet}, \citenamefont {LaPlante}, \citenamefont {Wan}, \citenamefont
  {Pritchett}, \citenamefont {Lee}, \citenamefont {Hamalainen}, \citenamefont
  {Moore}, \citenamefont {Kerr},\ and\ \citenamefont
  {Jones}}]{sacchet2015attention}%
  \BibitemOpen
  \bibfield  {author} {\bibinfo {author} {\bibfnamefont {M.~D.}\ \bibnamefont
  {Sacchet}}, \bibinfo {author} {\bibfnamefont {R.~A.}\ \bibnamefont
  {LaPlante}}, \bibinfo {author} {\bibfnamefont {Q.}~\bibnamefont {Wan}},
  \bibinfo {author} {\bibfnamefont {D.~L.}\ \bibnamefont {Pritchett}}, \bibinfo
  {author} {\bibfnamefont {A.~K.}\ \bibnamefont {Lee}}, \bibinfo {author}
  {\bibfnamefont {M.}~\bibnamefont {Hamalainen}}, \bibinfo {author}
  {\bibfnamefont {C.~I.}\ \bibnamefont {Moore}}, \bibinfo {author}
  {\bibfnamefont {C.~E.}\ \bibnamefont {Kerr}}, \ and\ \bibinfo {author}
  {\bibfnamefont {S.~R.}\ \bibnamefont {Jones}},\ }\href@noop {} {\bibfield
  {journal} {\bibinfo  {journal} {J Neurosci}\ }\textbf {\bibinfo {volume}
  {35}},\ \bibinfo {pages} {2074} (\bibinfo {year} {2015})}\BibitemShut
  {NoStop}%
\bibitem [{\citenamefont {von Stein}\ \emph {et~al.}(1999)\citenamefont {von
  Stein}, \citenamefont {Rappelsberger}, \citenamefont {Sarnthein},\ and\
  \citenamefont {Petsche}}]{doi:10.1093cercor9.2.137}%
  \BibitemOpen
  \bibfield  {author} {\bibinfo {author} {\bibfnamefont {A.}~\bibnamefont {von
  Stein}}, \bibinfo {author} {\bibfnamefont {P.}~\bibnamefont {Rappelsberger}},
  \bibinfo {author} {\bibfnamefont {J.}~\bibnamefont {Sarnthein}}, \ and\
  \bibinfo {author} {\bibfnamefont {H.}~\bibnamefont {Petsche}},\ }\href@noop
  {} {\bibfield  {journal} {\bibinfo  {journal} {Cerebral Cortex}\ }\textbf
  {\bibinfo {volume} {9}},\ \bibinfo {pages} {137} (\bibinfo {year}
  {1999})}\BibitemShut {NoStop}%
\bibitem [{\citenamefont {Kopell}\ \emph {et~al.}(2000)\citenamefont {Kopell},
  \citenamefont {Ermentrout}, \citenamefont {Whittington},\ and\ \citenamefont
  {Traub}}]{Kopell1867}%
  \BibitemOpen
  \bibfield  {author} {\bibinfo {author} {\bibfnamefont {N.}~\bibnamefont
  {Kopell}}, \bibinfo {author} {\bibfnamefont {G.~B.}\ \bibnamefont
  {Ermentrout}}, \bibinfo {author} {\bibfnamefont {M.~A.}\ \bibnamefont
  {Whittington}}, \ and\ \bibinfo {author} {\bibfnamefont {R.~D.}\ \bibnamefont
  {Traub}},\ }\href {\doibase 10.1073/pnas.97.4.1867} {\bibfield  {journal}
  {\bibinfo  {journal} {Proceedings of the National Academy of Sciences}\
  }\textbf {\bibinfo {volume} {97}},\ \bibinfo {pages} {1867} (\bibinfo {year}
  {2000})},\ \Eprint
  {http://arxiv.org/abs/http://www.pnas.org/content/97/4/1867.full.pdf}
  {http://www.pnas.org/content/97/4/1867.full.pdf} \BibitemShut {NoStop}%
\bibitem [{\citenamefont {Sherman}\ \emph {et~al.}(2016)\citenamefont
  {Sherman}, \citenamefont {Lee}, \citenamefont {Law}, \citenamefont {Haegens},
  \citenamefont {Thorn}, \citenamefont {Hamalainen}, \citenamefont {Moore},\
  and\ \citenamefont {Jones}}]{sherman2016neural}%
  \BibitemOpen
  \bibfield  {author} {\bibinfo {author} {\bibfnamefont {M.~A.}\ \bibnamefont
  {Sherman}}, \bibinfo {author} {\bibfnamefont {S.}~\bibnamefont {Lee}},
  \bibinfo {author} {\bibfnamefont {R.}~\bibnamefont {Law}}, \bibinfo {author}
  {\bibfnamefont {S.}~\bibnamefont {Haegens}}, \bibinfo {author} {\bibfnamefont
  {C.~A.}\ \bibnamefont {Thorn}}, \bibinfo {author} {\bibfnamefont {M.~S.}\
  \bibnamefont {Hamalainen}}, \bibinfo {author} {\bibfnamefont {C.~I.}\
  \bibnamefont {Moore}}, \ and\ \bibinfo {author} {\bibfnamefont {S.~R.}\
  \bibnamefont {Jones}},\ }\href@noop {} {\bibfield  {journal} {\bibinfo
  {journal} {Proc Natl Acad Sci U S A}\ }\textbf {\bibinfo {volume} {113}},\
  \bibinfo {pages} {E4885} (\bibinfo {year} {2016})}\BibitemShut {NoStop}%
\bibitem [{\citenamefont {Lee}\ and\ \citenamefont
  {Jones}(2013)}]{lee2013distinguishing}%
  \BibitemOpen
  \bibfield  {author} {\bibinfo {author} {\bibfnamefont {S.}~\bibnamefont
  {Lee}}\ and\ \bibinfo {author} {\bibfnamefont {S.~R.}\ \bibnamefont
  {Jones}},\ }\href@noop {} {\bibfield  {journal} {\bibinfo  {journal} {Front
  Hum Neurosci}\ }\textbf {\bibinfo {volume} {7}},\ \bibinfo {pages} {869}
  (\bibinfo {year} {2013})}\BibitemShut {NoStop}%
\bibitem [{\citenamefont {Muldoon}\ \emph {et~al.}(2016)\citenamefont
  {Muldoon}, \citenamefont {Pasqualetti}, \citenamefont {Gu}, \citenamefont
  {Cieslak}, \citenamefont {Grafton}, \citenamefont {Vettel},\ and\
  \citenamefont {Bassett}}]{muldoon2016stimulation}%
  \BibitemOpen
  \bibfield  {author} {\bibinfo {author} {\bibfnamefont {S.~F.}\ \bibnamefont
  {Muldoon}}, \bibinfo {author} {\bibfnamefont {F.}~\bibnamefont
  {Pasqualetti}}, \bibinfo {author} {\bibfnamefont {S.}~\bibnamefont {Gu}},
  \bibinfo {author} {\bibfnamefont {M.}~\bibnamefont {Cieslak}}, \bibinfo
  {author} {\bibfnamefont {S.~T.}\ \bibnamefont {Grafton}}, \bibinfo {author}
  {\bibfnamefont {J.~M.}\ \bibnamefont {Vettel}}, \ and\ \bibinfo {author}
  {\bibfnamefont {D.~S.}\ \bibnamefont {Bassett}},\ }\href@noop {} {\bibfield
  {journal} {\bibinfo  {journal} {PLoS Comput Biol}\ }\textbf {\bibinfo
  {volume} {12}},\ \bibinfo {pages} {e1005076} (\bibinfo {year}
  {2016})}\BibitemShut {NoStop}%
\bibitem [{\citenamefont {Pecora}\ and\ \citenamefont
  {Carroll}(1998)}]{PhysRevLett.80.2109}%
  \BibitemOpen
  \bibfield  {author} {\bibinfo {author} {\bibfnamefont {L.~M.}\ \bibnamefont
  {Pecora}}\ and\ \bibinfo {author} {\bibfnamefont {T.~L.}\ \bibnamefont
  {Carroll}},\ }\href {\doibase 10.1103/PhysRevLett.80.2109} {\bibfield
  {journal} {\bibinfo  {journal} {Phys. Rev. Lett.}\ }\textbf {\bibinfo
  {volume} {80}},\ \bibinfo {pages} {2109} (\bibinfo {year}
  {1998})}\BibitemShut {NoStop}%
\bibitem [{\citenamefont {Tang}\ \emph {et~al.}(2017)\citenamefont {Tang},
  \citenamefont {Giusti}, \citenamefont {Baum}, \citenamefont {Gu},
  \citenamefont {Pollock}, \citenamefont {Kahn}, \citenamefont {Roalf},
  \citenamefont {Moore}, \citenamefont {Ruparel}, \citenamefont {Gur},
  \citenamefont {Gur}, \citenamefont {Satterthwaite},\ and\ \citenamefont
  {Bassett}}]{tang2017developmental}%
  \BibitemOpen
  \bibfield  {author} {\bibinfo {author} {\bibfnamefont {E.}~\bibnamefont
  {Tang}}, \bibinfo {author} {\bibfnamefont {C.}~\bibnamefont {Giusti}},
  \bibinfo {author} {\bibfnamefont {G.~L.}\ \bibnamefont {Baum}}, \bibinfo
  {author} {\bibfnamefont {S.}~\bibnamefont {Gu}}, \bibinfo {author}
  {\bibfnamefont {E.}~\bibnamefont {Pollock}}, \bibinfo {author} {\bibfnamefont
  {A.~E.}\ \bibnamefont {Kahn}}, \bibinfo {author} {\bibfnamefont {D.~R.}\
  \bibnamefont {Roalf}}, \bibinfo {author} {\bibfnamefont {T.~M.}\ \bibnamefont
  {Moore}}, \bibinfo {author} {\bibfnamefont {K.}~\bibnamefont {Ruparel}},
  \bibinfo {author} {\bibfnamefont {R.~C.}\ \bibnamefont {Gur}}, \bibinfo
  {author} {\bibfnamefont {R.~E.}\ \bibnamefont {Gur}}, \bibinfo {author}
  {\bibfnamefont {T.~D.}\ \bibnamefont {Satterthwaite}}, \ and\ \bibinfo
  {author} {\bibfnamefont {D.~S.}\ \bibnamefont {Bassett}},\ }\href@noop {}
  {\bibfield  {journal} {\bibinfo  {journal} {Nat Commun}\ }\textbf {\bibinfo
  {volume} {8}},\ \bibinfo {pages} {1252} (\bibinfo {year} {2017})}\BibitemShut
  {NoStop}%
\bibitem [{\citenamefont {Tang}\ and\ \citenamefont
  {Bassett}(2018)}]{tang2018control}%
  \BibitemOpen
  \bibfield  {author} {\bibinfo {author} {\bibfnamefont {E.}~\bibnamefont
  {Tang}}\ and\ \bibinfo {author} {\bibfnamefont {D.~S.}\ \bibnamefont
  {Bassett}},\ }\href@noop {} {\bibfield  {journal} {\bibinfo  {journal} {Rev.
  Mod. Phys.}\ }\textbf {\bibinfo {volume} {90}},\ \bibinfo {pages} {031003}
  (\bibinfo {year} {2018})}\BibitemShut {NoStop}%
\bibitem [{\citenamefont {Gu}\ \emph {et~al.}(2015)\citenamefont {Gu},
  \citenamefont {Pasqualetti}, \citenamefont {Cieslak}, \citenamefont
  {Telesford}, \citenamefont {Alfred}, \citenamefont {Kahn}, \citenamefont
  {Medaglia}, \citenamefont {Vettel}, \citenamefont {Miller}, \citenamefont
  {Grafton} \emph {et~al.}}]{gu2015controllability}%
  \BibitemOpen
  \bibfield  {author} {\bibinfo {author} {\bibfnamefont {S.}~\bibnamefont
  {Gu}}, \bibinfo {author} {\bibfnamefont {F.}~\bibnamefont {Pasqualetti}},
  \bibinfo {author} {\bibfnamefont {M.}~\bibnamefont {Cieslak}}, \bibinfo
  {author} {\bibfnamefont {Q.~K.}\ \bibnamefont {Telesford}}, \bibinfo {author}
  {\bibfnamefont {B.~Y.}\ \bibnamefont {Alfred}}, \bibinfo {author}
  {\bibfnamefont {A.~E.}\ \bibnamefont {Kahn}}, \bibinfo {author}
  {\bibfnamefont {J.~D.}\ \bibnamefont {Medaglia}}, \bibinfo {author}
  {\bibfnamefont {J.~M.}\ \bibnamefont {Vettel}}, \bibinfo {author}
  {\bibfnamefont {M.~B.}\ \bibnamefont {Miller}}, \bibinfo {author}
  {\bibfnamefont {S.~T.}\ \bibnamefont {Grafton}},  \emph {et~al.},\
  }\href@noop {} {\bibfield  {journal} {\bibinfo  {journal} {Nature
  communications}\ }\textbf {\bibinfo {volume} {6}} (\bibinfo {year}
  {2015})}\BibitemShut {NoStop}%
\bibitem [{\citenamefont {Taylor}\ \emph {et~al.}(2015)\citenamefont {Taylor},
  \citenamefont {Thomas}, \citenamefont {Sinha}, \citenamefont {Dauwels},
  \citenamefont {Kaiser}, \citenamefont {Thesen},\ and\ \citenamefont
  {Ruths}}]{taylor2015optimal}%
  \BibitemOpen
  \bibfield  {author} {\bibinfo {author} {\bibfnamefont {P.~N.}\ \bibnamefont
  {Taylor}}, \bibinfo {author} {\bibfnamefont {J.}~\bibnamefont {Thomas}},
  \bibinfo {author} {\bibfnamefont {N.}~\bibnamefont {Sinha}}, \bibinfo
  {author} {\bibfnamefont {J.}~\bibnamefont {Dauwels}}, \bibinfo {author}
  {\bibfnamefont {M.}~\bibnamefont {Kaiser}}, \bibinfo {author} {\bibfnamefont
  {T.}~\bibnamefont {Thesen}}, \ and\ \bibinfo {author} {\bibfnamefont
  {J.}~\bibnamefont {Ruths}},\ }\href@noop {} {\bibfield  {journal} {\bibinfo
  {journal} {Front Neurosci}\ }\textbf {\bibinfo {volume} {9}},\ \bibinfo
  {pages} {202} (\bibinfo {year} {2015})}\BibitemShut {NoStop}%
\bibitem [{\citenamefont {Yan}\ \emph {et~al.}(2017{\natexlab{a}})\citenamefont
  {Yan}, \citenamefont {V{\'e}rtes}, \citenamefont {Towlson}, \citenamefont
  {Chew}, \citenamefont {Walker}, \citenamefont {Schafer},\ and\ \citenamefont
  {Barab{\'a}si}}]{Yan2017}%
  \BibitemOpen
  \bibfield  {author} {\bibinfo {author} {\bibfnamefont {G.}~\bibnamefont
  {Yan}}, \bibinfo {author} {\bibfnamefont {P.~E.}\ \bibnamefont {V{\'e}rtes}},
  \bibinfo {author} {\bibfnamefont {E.~K.}\ \bibnamefont {Towlson}}, \bibinfo
  {author} {\bibfnamefont {Y.~L.}\ \bibnamefont {Chew}}, \bibinfo {author}
  {\bibfnamefont {D.~S.}\ \bibnamefont {Walker}}, \bibinfo {author}
  {\bibfnamefont {W.~R.}\ \bibnamefont {Schafer}}, \ and\ \bibinfo {author}
  {\bibfnamefont {A.-L.}\ \bibnamefont {Barab{\'a}si}},\ }\href@noop {}
  {\bibfield  {journal} {\bibinfo  {journal} {Nature}\ }\textbf {\bibinfo
  {volume} {550}},\ \bibinfo {pages} {519–523} (\bibinfo {year}
  {2017}{\natexlab{a}})}\BibitemShut {NoStop}%
\bibitem [{\citenamefont {Kim}\ \emph {et~al.}(2018)\citenamefont {Kim},
  \citenamefont {Soffer}, \citenamefont {Kahn}, \citenamefont {Vettel},
  \citenamefont {Pasqualetti},\ and\ \citenamefont {Bassett}}]{kim2018role}%
  \BibitemOpen
  \bibfield  {author} {\bibinfo {author} {\bibfnamefont {J.~Z.}\ \bibnamefont
  {Kim}}, \bibinfo {author} {\bibfnamefont {J.~M.}\ \bibnamefont {Soffer}},
  \bibinfo {author} {\bibfnamefont {A.~E.}\ \bibnamefont {Kahn}}, \bibinfo
  {author} {\bibfnamefont {J.~M.}\ \bibnamefont {Vettel}}, \bibinfo {author}
  {\bibfnamefont {F.}~\bibnamefont {Pasqualetti}}, \ and\ \bibinfo {author}
  {\bibfnamefont {D.~S.}\ \bibnamefont {Bassett}},\ }\href@noop {} {\bibfield
  {journal} {\bibinfo  {journal} {Nat Phys}\ }\textbf {\bibinfo {volume}
  {14}},\ \bibinfo {pages} {91} (\bibinfo {year} {2018})}\BibitemShut {NoStop}%
\bibitem [{\citenamefont {Pasqualetti}\ \emph
  {et~al.}(2014{\natexlab{a}})\citenamefont {Pasqualetti}, \citenamefont
  {Zampieri},\ and\ \citenamefont {Bullo}}]{FP-SZ-FB:13q}%
  \BibitemOpen
  \bibfield  {author} {\bibinfo {author} {\bibfnamefont {F.}~\bibnamefont
  {Pasqualetti}}, \bibinfo {author} {\bibfnamefont {S.}~\bibnamefont
  {Zampieri}}, \ and\ \bibinfo {author} {\bibfnamefont {F.}~\bibnamefont
  {Bullo}},\ }\href@noop {} {\bibfield  {journal} {\bibinfo  {journal} {IEEE
  Transactions on Control of Network Systems}\ }\textbf {\bibinfo {volume}
  {1}},\ \bibinfo {pages} {40} (\bibinfo {year}
  {2014}{\natexlab{a}})}\BibitemShut {NoStop}%
\bibitem [{\citenamefont {Cornblath}\ \emph {et~al.}(2018)\citenamefont
  {Cornblath}, \citenamefont {Tang}, \citenamefont {Baum}, \citenamefont
  {Moore}, \citenamefont {Adebimpe}, \citenamefont {Roalf}, \citenamefont
  {Gur}, \citenamefont {Gur}, \citenamefont {Pasqualetti}, \citenamefont
  {Satterthwaite},\ and\ \citenamefont {Bassett}}]{cornblath2018sex}%
  \BibitemOpen
  \bibfield  {author} {\bibinfo {author} {\bibfnamefont {E.~J.}\ \bibnamefont
  {Cornblath}}, \bibinfo {author} {\bibfnamefont {E.}~\bibnamefont {Tang}},
  \bibinfo {author} {\bibfnamefont {G.~L.}\ \bibnamefont {Baum}}, \bibinfo
  {author} {\bibfnamefont {T.~M.}\ \bibnamefont {Moore}}, \bibinfo {author}
  {\bibfnamefont {A.}~\bibnamefont {Adebimpe}}, \bibinfo {author}
  {\bibfnamefont {D.~R.}\ \bibnamefont {Roalf}}, \bibinfo {author}
  {\bibfnamefont {R.~C.}\ \bibnamefont {Gur}}, \bibinfo {author} {\bibfnamefont
  {R.~E.}\ \bibnamefont {Gur}}, \bibinfo {author} {\bibfnamefont
  {F.}~\bibnamefont {Pasqualetti}}, \bibinfo {author} {\bibfnamefont {T.~D.}\
  \bibnamefont {Satterthwaite}}, \ and\ \bibinfo {author} {\bibfnamefont
  {D.~S.}\ \bibnamefont {Bassett}},\ }\href@noop {} {\bibfield  {journal}
  {\bibinfo  {journal} {Neuroimage}\ }\textbf {\bibinfo {volume} {188}},\
  \bibinfo {pages} {122} (\bibinfo {year} {2018})}\BibitemShut {NoStop}%
\bibitem [{\citenamefont {Jeganathan}\ \emph {et~al.}(2018)\citenamefont
  {Jeganathan}, \citenamefont {Perry}, \citenamefont {Bassett}, \citenamefont
  {Roberts}, \citenamefont {Mitchell},\ and\ \citenamefont
  {Breakspear}}]{jeganathan2018fronto}%
  \BibitemOpen
  \bibfield  {author} {\bibinfo {author} {\bibfnamefont {J.}~\bibnamefont
  {Jeganathan}}, \bibinfo {author} {\bibfnamefont {A.}~\bibnamefont {Perry}},
  \bibinfo {author} {\bibfnamefont {D.~S.}\ \bibnamefont {Bassett}}, \bibinfo
  {author} {\bibfnamefont {G.}~\bibnamefont {Roberts}}, \bibinfo {author}
  {\bibfnamefont {P.~B.}\ \bibnamefont {Mitchell}}, \ and\ \bibinfo {author}
  {\bibfnamefont {M.}~\bibnamefont {Breakspear}},\ }\href@noop {} {\bibfield
  {journal} {\bibinfo  {journal} {Neuroimage Clin}\ }\textbf {\bibinfo {volume}
  {19}},\ \bibinfo {pages} {71} (\bibinfo {year} {2018})}\BibitemShut {NoStop}%
\bibitem [{\citenamefont {Motter}(2015)}]{motter2015networkcontrology}%
  \BibitemOpen
  \bibfield  {author} {\bibinfo {author} {\bibfnamefont {A.~E.}\ \bibnamefont
  {Motter}},\ }\href@noop {} {\bibfield  {journal} {\bibinfo  {journal}
  {Chaos}\ }\textbf {\bibinfo {volume} {25}},\ \bibinfo {pages} {097621}
  (\bibinfo {year} {2015})}\BibitemShut {NoStop}%
\bibitem [{\citenamefont {Liu}\ and\ \citenamefont
  {Barab\'asi}(2016)}]{RevModPhys.88.035006}%
  \BibitemOpen
  \bibfield  {author} {\bibinfo {author} {\bibfnamefont {Y.-Y.}\ \bibnamefont
  {Liu}}\ and\ \bibinfo {author} {\bibfnamefont {A.-L.}\ \bibnamefont
  {Barab\'asi}},\ }\href {\doibase 10.1103/RevModPhys.88.035006} {\bibfield
  {journal} {\bibinfo  {journal} {Rev. Mod. Phys.}\ }\textbf {\bibinfo {volume}
  {88}},\ \bibinfo {pages} {035006} (\bibinfo {year} {2016})}\BibitemShut
  {NoStop}%
\bibitem [{\citenamefont {Satterthwaite}\ \emph {et~al.}(2016)\citenamefont
  {Satterthwaite}, \citenamefont {Connolly}, \citenamefont {Ruparel},
  \citenamefont {Calkins}, \citenamefont {Jackson}, \citenamefont {Elliott},
  \citenamefont {Roalf}, \citenamefont {Ryan~Hopsona}, \citenamefont {Behr},
  \citenamefont {Qiu}, \citenamefont {Mentch}, \citenamefont {Chiavacci},
  \citenamefont {Sleiman}, \citenamefont {Gur}, \citenamefont {Hakonarson},\
  and\ \citenamefont {Gur}}]{satterthwaite2016philadelphia}%
  \BibitemOpen
  \bibfield  {author} {\bibinfo {author} {\bibfnamefont {T.~D.}\ \bibnamefont
  {Satterthwaite}}, \bibinfo {author} {\bibfnamefont {J.~J.}\ \bibnamefont
  {Connolly}}, \bibinfo {author} {\bibfnamefont {K.}~\bibnamefont {Ruparel}},
  \bibinfo {author} {\bibfnamefont {M.~E.}\ \bibnamefont {Calkins}}, \bibinfo
  {author} {\bibfnamefont {C.}~\bibnamefont {Jackson}}, \bibinfo {author}
  {\bibfnamefont {M.~A.}\ \bibnamefont {Elliott}}, \bibinfo {author}
  {\bibfnamefont {D.~R.}\ \bibnamefont {Roalf}}, \bibinfo {author}
  {\bibfnamefont {K.~P.}\ \bibnamefont {Ryan~Hopsona}}, \bibinfo {author}
  {\bibfnamefont {M.}~\bibnamefont {Behr}}, \bibinfo {author} {\bibfnamefont
  {H.}~\bibnamefont {Qiu}}, \bibinfo {author} {\bibfnamefont {F.~D.}\
  \bibnamefont {Mentch}}, \bibinfo {author} {\bibfnamefont {R.}~\bibnamefont
  {Chiavacci}}, \bibinfo {author} {\bibfnamefont {P.~M.}\ \bibnamefont
  {Sleiman}}, \bibinfo {author} {\bibfnamefont {R.~C.}\ \bibnamefont {Gur}},
  \bibinfo {author} {\bibfnamefont {H.}~\bibnamefont {Hakonarson}}, \ and\
  \bibinfo {author} {\bibfnamefont {R.~E.}\ \bibnamefont {Gur}},\ }\href@noop
  {} {\bibfield  {journal} {\bibinfo  {journal} {Neuroimage}\ }\textbf
  {\bibinfo {volume} {124}},\ \bibinfo {pages} {1115} (\bibinfo {year}
  {2016})}\BibitemShut {NoStop}%
\bibitem [{\citenamefont {Calkins}\ \emph {et~al.}(2015)\citenamefont
  {Calkins}, \citenamefont {Merikangas}, \citenamefont {Moore}, \citenamefont
  {Burstein}, \citenamefont {Behr}, \citenamefont {Satterthwaite},
  \citenamefont {Ruparel}, \citenamefont {Wolf}, \citenamefont {Roalf},
  \citenamefont {Mentch}, \citenamefont {Qiu}, \citenamefont {Chiavacci},
  \citenamefont {Connolly}, \citenamefont {Sleiman}, \citenamefont {Gur},
  \citenamefont {Hakonarson},\ and\ \citenamefont
  {Gur}}]{calkins2015philadelphia}%
  \BibitemOpen
  \bibfield  {author} {\bibinfo {author} {\bibfnamefont {M.~E.}\ \bibnamefont
  {Calkins}}, \bibinfo {author} {\bibfnamefont {K.~R.}\ \bibnamefont
  {Merikangas}}, \bibinfo {author} {\bibfnamefont {T.~M.}\ \bibnamefont
  {Moore}}, \bibinfo {author} {\bibfnamefont {M.}~\bibnamefont {Burstein}},
  \bibinfo {author} {\bibfnamefont {M.~A.}\ \bibnamefont {Behr}}, \bibinfo
  {author} {\bibfnamefont {T.~D.}\ \bibnamefont {Satterthwaite}}, \bibinfo
  {author} {\bibfnamefont {K.}~\bibnamefont {Ruparel}}, \bibinfo {author}
  {\bibfnamefont {D.~H.}\ \bibnamefont {Wolf}}, \bibinfo {author}
  {\bibfnamefont {D.~R.}\ \bibnamefont {Roalf}}, \bibinfo {author}
  {\bibfnamefont {F.~D.}\ \bibnamefont {Mentch}}, \bibinfo {author}
  {\bibfnamefont {H.}~\bibnamefont {Qiu}}, \bibinfo {author} {\bibfnamefont
  {R.}~\bibnamefont {Chiavacci}}, \bibinfo {author} {\bibfnamefont {J.~J.}\
  \bibnamefont {Connolly}}, \bibinfo {author} {\bibfnamefont {P.~M.~A.}\
  \bibnamefont {Sleiman}}, \bibinfo {author} {\bibfnamefont {R.~C.}\
  \bibnamefont {Gur}}, \bibinfo {author} {\bibfnamefont {H.}~\bibnamefont
  {Hakonarson}}, \ and\ \bibinfo {author} {\bibfnamefont {R.~E.}\ \bibnamefont
  {Gur}},\ }\href@noop {} {\bibfield  {journal} {\bibinfo  {journal} {J Child
  Psychol Psychiatry}\ }\textbf {\bibinfo {volume} {56}},\ \bibinfo {pages}
  {1356} (\bibinfo {year} {2015})}\BibitemShut {NoStop}%
\bibitem [{\citenamefont {Atasoy}\ \emph {et~al.}(2016)\citenamefont {Atasoy},
  \citenamefont {Donnelly},\ and\ \citenamefont {Pearson}}]{Atasoy2016}%
  \BibitemOpen
  \bibfield  {author} {\bibinfo {author} {\bibfnamefont {S.}~\bibnamefont
  {Atasoy}}, \bibinfo {author} {\bibfnamefont {I.}~\bibnamefont {Donnelly}}, \
  and\ \bibinfo {author} {\bibfnamefont {J.}~\bibnamefont {Pearson}},\
  }\href@noop {} {\bibfield  {journal} {\bibinfo  {journal} {Nature
  Communications}\ }\textbf {\bibinfo {volume} {7}},\ \bibinfo {pages} {10340}
  (\bibinfo {year} {2016})},\ \bibinfo {note} {article}\BibitemShut {NoStop}%
\bibitem [{\citenamefont {He}\ \emph {et~al.}(2014)\citenamefont {He},
  \citenamefont {Wang}, \citenamefont {Zhang},\ and\ \citenamefont
  {Zhan}}]{PhysRevE.90.012909}%
  \BibitemOpen
  \bibfield  {author} {\bibinfo {author} {\bibfnamefont {Z.}~\bibnamefont
  {He}}, \bibinfo {author} {\bibfnamefont {X.}~\bibnamefont {Wang}}, \bibinfo
  {author} {\bibfnamefont {G.-Y.}\ \bibnamefont {Zhang}}, \ and\ \bibinfo
  {author} {\bibfnamefont {M.}~\bibnamefont {Zhan}},\ }\href {\doibase
  10.1103/PhysRevE.90.012909} {\bibfield  {journal} {\bibinfo  {journal} {Phys.
  Rev. E}\ }\textbf {\bibinfo {volume} {90}},\ \bibinfo {pages} {012909}
  (\bibinfo {year} {2014})}\BibitemShut {NoStop}%
\bibitem [{\citenamefont {Baum}\ \emph {et~al.}(2017)\citenamefont {Baum},
  \citenamefont {Ciric}, \citenamefont {Roalf}, \citenamefont {Betzel},
  \citenamefont {Moore}, \citenamefont {Shinohara}, \citenamefont {Kahn},
  \citenamefont {Vandekar}, \citenamefont {Rupert}, \citenamefont {Quarmley},
  \citenamefont {Cook}, \citenamefont {Elliott}, \citenamefont {Ruparel},
  \citenamefont {Gur}, \citenamefont {Gur}, \citenamefont {Bassett},\ and\
  \citenamefont {Satterthwaite}}]{baum2017modular}%
  \BibitemOpen
  \bibfield  {author} {\bibinfo {author} {\bibfnamefont {G.~L.}\ \bibnamefont
  {Baum}}, \bibinfo {author} {\bibfnamefont {R.}~\bibnamefont {Ciric}},
  \bibinfo {author} {\bibfnamefont {D.~R.}\ \bibnamefont {Roalf}}, \bibinfo
  {author} {\bibfnamefont {R.~F.}\ \bibnamefont {Betzel}}, \bibinfo {author}
  {\bibfnamefont {T.~M.}\ \bibnamefont {Moore}}, \bibinfo {author}
  {\bibfnamefont {R.~T.}\ \bibnamefont {Shinohara}}, \bibinfo {author}
  {\bibfnamefont {A.~E.}\ \bibnamefont {Kahn}}, \bibinfo {author}
  {\bibfnamefont {S.~N.}\ \bibnamefont {Vandekar}}, \bibinfo {author}
  {\bibfnamefont {P.~E.}\ \bibnamefont {Rupert}}, \bibinfo {author}
  {\bibfnamefont {M.}~\bibnamefont {Quarmley}}, \bibinfo {author}
  {\bibfnamefont {P.~A.}\ \bibnamefont {Cook}}, \bibinfo {author}
  {\bibfnamefont {M.~A.}\ \bibnamefont {Elliott}}, \bibinfo {author}
  {\bibfnamefont {K.}~\bibnamefont {Ruparel}}, \bibinfo {author} {\bibfnamefont
  {R.~E.}\ \bibnamefont {Gur}}, \bibinfo {author} {\bibfnamefont {R.~C.}\
  \bibnamefont {Gur}}, \bibinfo {author} {\bibfnamefont {D.~S.}\ \bibnamefont
  {Bassett}}, \ and\ \bibinfo {author} {\bibfnamefont {T.~D.}\ \bibnamefont
  {Satterthwaite}},\ }\href@noop {} {\bibfield  {journal} {\bibinfo  {journal}
  {Curr Biol}\ }\textbf {\bibinfo {volume} {27}},\ \bibinfo {pages} {1561}
  (\bibinfo {year} {2017})}\BibitemShut {NoStop}%
\bibitem [{\citenamefont {Baum}\ \emph {et~al.}(2018)\citenamefont {Baum},
  \citenamefont {Roalf}, \citenamefont {Cook}, \citenamefont {Ciric},
  \citenamefont {Rosen}, \citenamefont {Xia}, \citenamefont {Elliott},
  \citenamefont {Ruparel}, \citenamefont {Verma}, \citenamefont {Tunc},
  \citenamefont {Gur}, \citenamefont {Gur}, \citenamefont {Bassett},\ and\
  \citenamefont {Satterthwaite}}]{baum2018impact}%
  \BibitemOpen
  \bibfield  {author} {\bibinfo {author} {\bibfnamefont {G.~L.}\ \bibnamefont
  {Baum}}, \bibinfo {author} {\bibfnamefont {D.~R.}\ \bibnamefont {Roalf}},
  \bibinfo {author} {\bibfnamefont {P.~A.}\ \bibnamefont {Cook}}, \bibinfo
  {author} {\bibfnamefont {R.}~\bibnamefont {Ciric}}, \bibinfo {author}
  {\bibfnamefont {A.~F.~G.}\ \bibnamefont {Rosen}}, \bibinfo {author}
  {\bibfnamefont {C.}~\bibnamefont {Xia}}, \bibinfo {author} {\bibfnamefont
  {M.~A.}\ \bibnamefont {Elliott}}, \bibinfo {author} {\bibfnamefont
  {K.}~\bibnamefont {Ruparel}}, \bibinfo {author} {\bibfnamefont
  {R.}~\bibnamefont {Verma}}, \bibinfo {author} {\bibfnamefont
  {B.}~\bibnamefont {Tunc}}, \bibinfo {author} {\bibfnamefont {R.~C.}\
  \bibnamefont {Gur}}, \bibinfo {author} {\bibfnamefont {R.~E.}\ \bibnamefont
  {Gur}}, \bibinfo {author} {\bibfnamefont {D.~S.}\ \bibnamefont {Bassett}}, \
  and\ \bibinfo {author} {\bibfnamefont {T.~D.}\ \bibnamefont
  {Satterthwaite}},\ }\href@noop {} {\bibfield  {journal} {\bibinfo  {journal}
  {Neuroimage}\ }\textbf {\bibinfo {volume} {173}},\ \bibinfo {pages} {275}
  (\bibinfo {year} {2018})}\BibitemShut {NoStop}%
\bibitem [{\citenamefont {Kalman}\ \emph {et~al.}(1963)\citenamefont {Kalman},
  \citenamefont {Ho},\ and\ \citenamefont {Narendra}}]{rek-ych-skn:63_2}%
  \BibitemOpen
  \bibfield  {author} {\bibinfo {author} {\bibfnamefont {R.~E.}\ \bibnamefont
  {Kalman}}, \bibinfo {author} {\bibfnamefont {Y.~C.}\ \bibnamefont {Ho}}, \
  and\ \bibinfo {author} {\bibfnamefont {K.~S.}\ \bibnamefont {Narendra}},\
  }\href@noop {} {\bibfield  {journal} {\bibinfo  {journal} {Contributions to
  Differential Equations}\ }\textbf {\bibinfo {volume} {1}},\ \bibinfo {pages}
  {189} (\bibinfo {year} {1963})}\BibitemShut {NoStop}%
\bibitem [{\citenamefont {Kailath}(1980)}]{kailath1980linear}%
  \BibitemOpen
  \bibfield  {author} {\bibinfo {author} {\bibfnamefont {T.}~\bibnamefont
  {Kailath}},\ }\href@noop {} {\emph {\bibinfo {title} {Linear systems}}},\
  Vol.~\bibinfo {volume} {1}\ (\bibinfo  {publisher} {Prentice-Hall Englewood
  Cliffs, NJ},\ \bibinfo {year} {1980})\BibitemShut {NoStop}%
\bibitem [{\citenamefont {Hewer}\ and\ \citenamefont
  {Kenney}(1988)}]{hewer1988sensitivity}%
  \BibitemOpen
  \bibfield  {author} {\bibinfo {author} {\bibfnamefont {G.}~\bibnamefont
  {Hewer}}\ and\ \bibinfo {author} {\bibfnamefont {C.}~\bibnamefont {Kenney}},\
  }\href@noop {} {\bibfield  {journal} {\bibinfo  {journal} {SIAM journal on
  control and optimization}\ }\textbf {\bibinfo {volume} {26}},\ \bibinfo
  {pages} {321} (\bibinfo {year} {1988})}\BibitemShut {NoStop}%
\bibitem [{\citenamefont {Gahinet}\ \emph {et~al.}(1990)\citenamefont
  {Gahinet}, \citenamefont {Laub}, \citenamefont {Kenney},\ and\ \citenamefont
  {Hewer}}]{gahinet1990sensitivity}%
  \BibitemOpen
  \bibfield  {author} {\bibinfo {author} {\bibfnamefont {P.~M.}\ \bibnamefont
  {Gahinet}}, \bibinfo {author} {\bibfnamefont {A.~J.}\ \bibnamefont {Laub}},
  \bibinfo {author} {\bibfnamefont {C.~S.}\ \bibnamefont {Kenney}}, \ and\
  \bibinfo {author} {\bibfnamefont {G.~A.}\ \bibnamefont {Hewer}},\ }\href@noop
  {} {\bibfield  {journal} {\bibinfo  {journal} {IEEE Transactions on Automatic
  Control}\ }\textbf {\bibinfo {volume} {35}},\ \bibinfo {pages} {1209}
  (\bibinfo {year} {1990})}\BibitemShut {NoStop}%
\bibitem [{\citenamefont {Hammarling}(1982)}]{hammarling1982numerical}%
  \BibitemOpen
  \bibfield  {author} {\bibinfo {author} {\bibfnamefont {S.~J.}\ \bibnamefont
  {Hammarling}},\ }\href@noop {} {\bibfield  {journal} {\bibinfo  {journal}
  {IMA Journal of Numerical Analysis}\ }\textbf {\bibinfo {volume} {2}},\
  \bibinfo {pages} {303} (\bibinfo {year} {1982})}\BibitemShut {NoStop}%
\bibitem [{\citenamefont {Sorensen}\ and\ \citenamefont
  {Zhou}(2002)}]{sorensen2002bounds}%
  \BibitemOpen
  \bibfield  {author} {\bibinfo {author} {\bibfnamefont {D.~C.}\ \bibnamefont
  {Sorensen}}\ and\ \bibinfo {author} {\bibfnamefont {Y.}~\bibnamefont
  {Zhou}},\ }\href@noop {} {\emph {\bibinfo {title} {Bounds on eigenvalue decay
  rates and sensitivity of solutions to Lyapunov equations}}},\ \bibinfo {type}
  {Tech. Rep.}\ (\bibinfo {year} {2002})\BibitemShut {NoStop}%
\bibitem [{\citenamefont {Pasqualetti}\ \emph
  {et~al.}(2014{\natexlab{b}})\citenamefont {Pasqualetti}, \citenamefont
  {Zampieri},\ and\ \citenamefont {Bullo}}]{pasqualetti2014controllability}%
  \BibitemOpen
  \bibfield  {author} {\bibinfo {author} {\bibfnamefont {F.}~\bibnamefont
  {Pasqualetti}}, \bibinfo {author} {\bibfnamefont {S.}~\bibnamefont
  {Zampieri}}, \ and\ \bibinfo {author} {\bibfnamefont {F.}~\bibnamefont
  {Bullo}},\ }\href@noop {} {\bibfield  {journal} {\bibinfo  {journal} {Control
  of Network Systems, IEEE Transactions on}\ }\textbf {\bibinfo {volume} {1}},\
  \bibinfo {pages} {40} (\bibinfo {year} {2014}{\natexlab{b}})}\BibitemShut
  {NoStop}%
\bibitem [{\citenamefont {Wu-Yan}\ \emph {et~al.}(2018)\citenamefont {Wu-Yan},
  \citenamefont {Betzel}, \citenamefont {Tang}, \citenamefont {Gu},
  \citenamefont {Pasqualetti},\ and\ \citenamefont
  {Bassett}}]{wuyan2018benchmarking}%
  \BibitemOpen
  \bibfield  {author} {\bibinfo {author} {\bibfnamefont {E.}~\bibnamefont
  {Wu-Yan}}, \bibinfo {author} {\bibfnamefont {R.~F.}\ \bibnamefont {Betzel}},
  \bibinfo {author} {\bibfnamefont {E.}~\bibnamefont {Tang}}, \bibinfo {author}
  {\bibfnamefont {S.}~\bibnamefont {Gu}}, \bibinfo {author} {\bibfnamefont
  {F.}~\bibnamefont {Pasqualetti}}, \ and\ \bibinfo {author} {\bibfnamefont
  {D.~S.}\ \bibnamefont {Bassett}},\ }\href@noop {} {\bibfield  {journal}
  {\bibinfo  {journal} {J Nonlinear Sci}\ }\textbf {\bibinfo {volume} {Epub
  Ahead of Print}},\ \bibinfo {pages} {1} (\bibinfo {year} {2018})}\BibitemShut
  {NoStop}%
\bibitem [{\citenamefont {Roberts}\ \emph {et~al.}(2019)\citenamefont
  {Roberts}, \citenamefont {Gollo}, \citenamefont {Abeysuriya}, \citenamefont
  {Roberts}, \citenamefont {Mitchell}, \citenamefont {Woolrich},\ and\
  \citenamefont {Breakspear}}]{Roberts2019}%
  \BibitemOpen
  \bibfield  {author} {\bibinfo {author} {\bibfnamefont {J.~A.}\ \bibnamefont
  {Roberts}}, \bibinfo {author} {\bibfnamefont {L.~L.}\ \bibnamefont {Gollo}},
  \bibinfo {author} {\bibfnamefont {R.~G.}\ \bibnamefont {Abeysuriya}},
  \bibinfo {author} {\bibfnamefont {G.}~\bibnamefont {Roberts}}, \bibinfo
  {author} {\bibfnamefont {P.~B.}\ \bibnamefont {Mitchell}}, \bibinfo {author}
  {\bibfnamefont {M.~W.}\ \bibnamefont {Woolrich}}, \ and\ \bibinfo {author}
  {\bibfnamefont {M.}~\bibnamefont {Breakspear}},\ }\href {\doibase
  10.1038/s41467-019-08999-0} {\bibfield  {journal} {\bibinfo  {journal}
  {Nature Communications}\ }\textbf {\bibinfo {volume} {10}},\ \bibinfo {pages}
  {1056} (\bibinfo {year} {2019})}\BibitemShut {NoStop}%
\bibitem [{\citenamefont {Bassett}\ and\ \citenamefont
  {Bullmore}(2016)}]{bassett2016small}%
  \BibitemOpen
  \bibfield  {author} {\bibinfo {author} {\bibfnamefont {D.~S.}\ \bibnamefont
  {Bassett}}\ and\ \bibinfo {author} {\bibfnamefont {E.~T.}\ \bibnamefont
  {Bullmore}},\ }\href@noop {} {\bibfield  {journal} {\bibinfo  {journal}
  {Neuroscientist}\ }\textbf {\bibinfo {volume} {Epub ahead of print}}
  (\bibinfo {year} {2016})}\BibitemShut {NoStop}%
\bibitem [{\citenamefont {Bassett}\ and\ \citenamefont
  {Bullmore}(2006)}]{bassett2006small}%
  \BibitemOpen
  \bibfield  {author} {\bibinfo {author} {\bibfnamefont {D.~S.}\ \bibnamefont
  {Bassett}}\ and\ \bibinfo {author} {\bibfnamefont {E.}~\bibnamefont
  {Bullmore}},\ }\href@noop {} {\bibfield  {journal} {\bibinfo  {journal}
  {Neuroscientist}\ }\textbf {\bibinfo {volume} {12}},\ \bibinfo {pages} {512}
  (\bibinfo {year} {2006})}\BibitemShut {NoStop}%
\bibitem [{\citenamefont {Betzel}\ \emph {et~al.}(2018)\citenamefont {Betzel},
  \citenamefont {Medaglia},\ and\ \citenamefont
  {Bassett}}]{betzel2018diversity}%
  \BibitemOpen
  \bibfield  {author} {\bibinfo {author} {\bibfnamefont {R.~F.}\ \bibnamefont
  {Betzel}}, \bibinfo {author} {\bibfnamefont {J.~D.}\ \bibnamefont
  {Medaglia}}, \ and\ \bibinfo {author} {\bibfnamefont {D.~S.}\ \bibnamefont
  {Bassett}},\ }\href@noop {} {\bibfield  {journal} {\bibinfo  {journal} {Nat
  Commun}\ }\textbf {\bibinfo {volume} {9}},\ \bibinfo {pages} {346} (\bibinfo
  {year} {2018})}\BibitemShut {NoStop}%
\bibitem [{\citenamefont {Albert}\ and\ \citenamefont
  {Barabasi}(2002)}]{albert2002statistical}%
  \BibitemOpen
  \bibfield  {author} {\bibinfo {author} {\bibfnamefont {R.}~\bibnamefont
  {Albert}}\ and\ \bibinfo {author} {\bibfnamefont {A.-L.}\ \bibnamefont
  {Barabasi}},\ }\href@noop {} {\bibfield  {journal} {\bibinfo  {journal}
  {Reviews of Modern Physics}\ }\textbf {\bibinfo {volume} {74}},\ \bibinfo
  {pages} {47} (\bibinfo {year} {2002})}\BibitemShut {NoStop}%
\bibitem [{\citenamefont {Price}(1976)}]{price1976general}%
  \BibitemOpen
  \bibfield  {author} {\bibinfo {author} {\bibfnamefont {D.~J. d.~S.}\
  \bibnamefont {Price}},\ }\href@noop {} {\bibfield  {journal} {\bibinfo
  {journal} {Journal of the American Society for Information Science}\ }\textbf
  {\bibinfo {volume} {27}},\ \bibinfo {pages} {292} (\bibinfo {year}
  {1976})}\BibitemShut {NoStop}%
\bibitem [{\citenamefont {Barabasi}\ and\ \citenamefont
  {Albert}(1999)}]{barabasi1999emergence}%
  \BibitemOpen
  \bibfield  {author} {\bibinfo {author} {\bibfnamefont {A.~L.}\ \bibnamefont
  {Barabasi}}\ and\ \bibinfo {author} {\bibfnamefont {R.}~\bibnamefont
  {Albert}},\ }\href@noop {} {\bibfield  {journal} {\bibinfo  {journal}
  {Science}\ }\textbf {\bibinfo {volume} {286}},\ \bibinfo {pages} {509}
  (\bibinfo {year} {1999})}\BibitemShut {NoStop}%
\bibitem [{\citenamefont {Porfiri}(2011)}]{Porfiri_2011}%
  \BibitemOpen
  \bibfield  {author} {\bibinfo {author} {\bibfnamefont {M.}~\bibnamefont
  {Porfiri}},\ }\href {\doibase 10.1209/0295-5075/96/40014} {\bibfield
  {journal} {\bibinfo  {journal} {{EPL} (Europhysics Letters)}\ }\textbf
  {\bibinfo {volume} {96}},\ \bibinfo {pages} {40014} (\bibinfo {year}
  {2011})}\BibitemShut {NoStop}%
\bibitem [{\citenamefont {Sun}\ \emph {et~al.}(2009)\citenamefont {Sun},
  \citenamefont {Bollt},\ and\ \citenamefont {Nishikawa}}]{Sun_2009}%
  \BibitemOpen
  \bibfield  {author} {\bibinfo {author} {\bibfnamefont {J.}~\bibnamefont
  {Sun}}, \bibinfo {author} {\bibfnamefont {E.~M.}\ \bibnamefont {Bollt}}, \
  and\ \bibinfo {author} {\bibfnamefont {T.}~\bibnamefont {Nishikawa}},\ }\href
  {\doibase 10.1209/0295-5075/85/60011} {\bibfield  {journal} {\bibinfo
  {journal} {{EPL} (Europhysics Letters)}\ }\textbf {\bibinfo {volume} {85}},\
  \bibinfo {pages} {60011} (\bibinfo {year} {2009})}\BibitemShut {NoStop}%
\bibitem [{\citenamefont {Jiang}\ and\ \citenamefont {Lai}(2019)}]{Jiang2019}%
  \BibitemOpen
  \bibfield  {author} {\bibinfo {author} {\bibfnamefont {J.}~\bibnamefont
  {Jiang}}\ and\ \bibinfo {author} {\bibfnamefont {Y.-C.}\ \bibnamefont
  {Lai}},\ }\href {\doibase 10.1038/s41467-019-11822-5} {\bibfield  {journal}
  {\bibinfo  {journal} {Nature Communications}\ }\textbf {\bibinfo {volume}
  {10}},\ \bibinfo {pages} {3961} (\bibinfo {year} {2019})}\BibitemShut
  {NoStop}%
\bibitem [{\citenamefont {Huang}\ \emph {et~al.}(2009)\citenamefont {Huang},
  \citenamefont {Chen}, \citenamefont {Lai},\ and\ \citenamefont
  {Pecora}}]{PhysRevE.80.036204}%
  \BibitemOpen
  \bibfield  {author} {\bibinfo {author} {\bibfnamefont {L.}~\bibnamefont
  {Huang}}, \bibinfo {author} {\bibfnamefont {Q.}~\bibnamefont {Chen}},
  \bibinfo {author} {\bibfnamefont {Y.-C.}\ \bibnamefont {Lai}}, \ and\
  \bibinfo {author} {\bibfnamefont {L.~M.}\ \bibnamefont {Pecora}},\ }\href
  {\doibase 10.1103/PhysRevE.80.036204} {\bibfield  {journal} {\bibinfo
  {journal} {Phys. Rev. E}\ }\textbf {\bibinfo {volume} {80}},\ \bibinfo
  {pages} {036204} (\bibinfo {year} {2009})}\BibitemShut {NoStop}%
\bibitem [{\citenamefont {Tu}\ \emph {et~al.}(2018)\citenamefont {Tu},
  \citenamefont {Rocha}, \citenamefont {Corbetta}, \citenamefont {Zampieri},
  \citenamefont {Zorzi},\ and\ \citenamefont {Suweis}}]{tu2018warnings}%
  \BibitemOpen
  \bibfield  {author} {\bibinfo {author} {\bibfnamefont {C.}~\bibnamefont
  {Tu}}, \bibinfo {author} {\bibfnamefont {R.~P.}\ \bibnamefont {Rocha}},
  \bibinfo {author} {\bibfnamefont {M.}~\bibnamefont {Corbetta}}, \bibinfo
  {author} {\bibfnamefont {S.}~\bibnamefont {Zampieri}}, \bibinfo {author}
  {\bibfnamefont {M.}~\bibnamefont {Zorzi}}, \ and\ \bibinfo {author}
  {\bibfnamefont {S.}~\bibnamefont {Suweis}},\ }\href@noop {} {\bibfield
  {journal} {\bibinfo  {journal} {Neuroimage}\ }\textbf {\bibinfo {volume}
  {176}},\ \bibinfo {pages} {83} (\bibinfo {year} {2018})}\BibitemShut
  {NoStop}%
\bibitem [{\citenamefont {{RE: W}arnings}\ and\ \citenamefont {caveats in~brain
  controllability}(2019)}]{pasqualetti2019warnings}%
  \BibitemOpen
  \bibfield  {author} {\bibinfo {author} {\bibnamefont {{RE: W}arnings}}\ and\
  \bibinfo {author} {\bibnamefont {caveats in~brain controllability}},\
  }\href@noop {} {\bibfield  {journal} {\bibinfo  {journal} {Neuroimage}\
  }\textbf {\bibinfo {volume} {197}},\ \bibinfo {pages} {586} (\bibinfo {year}
  {2019})}\BibitemShut {NoStop}%
\bibitem [{\citenamefont {Menara}\ \emph {et~al.}(2017)\citenamefont {Menara},
  \citenamefont {Gu}, \citenamefont {Bassett},\ and\ \citenamefont
  {Pasqualetti}}]{menara2017structural}%
  \BibitemOpen
  \bibfield  {author} {\bibinfo {author} {\bibfnamefont {T.}~\bibnamefont
  {Menara}}, \bibinfo {author} {\bibfnamefont {S.}~\bibnamefont {Gu}}, \bibinfo
  {author} {\bibfnamefont {D.~S.}\ \bibnamefont {Bassett}}, \ and\ \bibinfo
  {author} {\bibfnamefont {F.}~\bibnamefont {Pasqualetti}},\ }\href@noop {}
  {\bibfield  {journal} {\bibinfo  {journal} {arXiv}\ }\textbf {\bibinfo
  {volume} {1706}},\ \bibinfo {pages} {05120} (\bibinfo {year}
  {2017})}\BibitemShut {NoStop}%
\bibitem [{\citenamefont {Yan}\ \emph {et~al.}(2017{\natexlab{b}})\citenamefont
  {Yan}, \citenamefont {Vértes}, \citenamefont {Towlson}, \citenamefont
  {Chew}, \citenamefont {Walker}, \citenamefont {Schafer},\ and\ \citenamefont
  {Barabási}}]{yan2017network}%
  \BibitemOpen
  \bibfield  {author} {\bibinfo {author} {\bibfnamefont {G.}~\bibnamefont
  {Yan}}, \bibinfo {author} {\bibfnamefont {P.~E.}\ \bibnamefont {Vértes}},
  \bibinfo {author} {\bibfnamefont {E.~K.}\ \bibnamefont {Towlson}}, \bibinfo
  {author} {\bibfnamefont {Y.~L.}\ \bibnamefont {Chew}}, \bibinfo {author}
  {\bibfnamefont {D.~S.}\ \bibnamefont {Walker}}, \bibinfo {author}
  {\bibfnamefont {W.~R.}\ \bibnamefont {Schafer}}, \ and\ \bibinfo {author}
  {\bibfnamefont {A.~L.}\ \bibnamefont {Barabási}},\ }\href@noop {} {\bibfield
   {journal} {\bibinfo  {journal} {Nature}\ }\textbf {\bibinfo {volume}
  {550}},\ \bibinfo {pages} {519} (\bibinfo {year}
  {2017}{\natexlab{b}})}\BibitemShut {NoStop}%
\bibitem [{\citenamefont {Zañudo}\ \emph {et~al.}(2017)\citenamefont
  {Zañudo}, \citenamefont {Yang},\ and\ \citenamefont
  {Albert}}]{zanudo2017structure}%
  \BibitemOpen
  \bibfield  {author} {\bibinfo {author} {\bibfnamefont {J.~G.~T.}\
  \bibnamefont {Zañudo}}, \bibinfo {author} {\bibfnamefont {G.}~\bibnamefont
  {Yang}}, \ and\ \bibinfo {author} {\bibfnamefont {R.}~\bibnamefont
  {Albert}},\ }\href@noop {} {\bibfield  {journal} {\bibinfo  {journal} {Proc
  Natl Acad Sci U S A}\ }\textbf {\bibinfo {volume} {114}},\ \bibinfo {pages}
  {7234} (\bibinfo {year} {2017})}\BibitemShut {NoStop}%
\bibitem [{\citenamefont {Gu}\ \emph {et~al.}(2017)\citenamefont {Gu},
  \citenamefont {Betzel}, \citenamefont {Mattar}, \citenamefont {Cieslak},
  \citenamefont {Delio}, \citenamefont {Grafton}, \citenamefont {Pasqualetti},\
  and\ \citenamefont {Bassett}}]{gu2017optimal}%
  \BibitemOpen
  \bibfield  {author} {\bibinfo {author} {\bibfnamefont {S.}~\bibnamefont
  {Gu}}, \bibinfo {author} {\bibfnamefont {R.~F.}\ \bibnamefont {Betzel}},
  \bibinfo {author} {\bibfnamefont {M.~G.}\ \bibnamefont {Mattar}}, \bibinfo
  {author} {\bibfnamefont {M.}~\bibnamefont {Cieslak}}, \bibinfo {author}
  {\bibfnamefont {P.~R.}\ \bibnamefont {Delio}}, \bibinfo {author}
  {\bibfnamefont {S.~T.}\ \bibnamefont {Grafton}}, \bibinfo {author}
  {\bibfnamefont {F.}~\bibnamefont {Pasqualetti}}, \ and\ \bibinfo {author}
  {\bibfnamefont {D.~S.}\ \bibnamefont {Bassett}},\ }\href@noop {} {\bibfield
  {journal} {\bibinfo  {journal} {Neuroimage}\ }\textbf {\bibinfo {volume}
  {148}},\ \bibinfo {pages} {305} (\bibinfo {year} {2017})}\BibitemShut
  {NoStop}%
\bibitem [{\citenamefont {Betzel}\ \emph {et~al.}(2016)\citenamefont {Betzel},
  \citenamefont {Gu}, \citenamefont {Medaglia}, \citenamefont {Pasqualetti},\
  and\ \citenamefont {Bassett}}]{betzel2016optimally}%
  \BibitemOpen
  \bibfield  {author} {\bibinfo {author} {\bibfnamefont {R.~F.}\ \bibnamefont
  {Betzel}}, \bibinfo {author} {\bibfnamefont {S.}~\bibnamefont {Gu}}, \bibinfo
  {author} {\bibfnamefont {J.~D.}\ \bibnamefont {Medaglia}}, \bibinfo {author}
  {\bibfnamefont {F.}~\bibnamefont {Pasqualetti}}, \ and\ \bibinfo {author}
  {\bibfnamefont {D.~S.}\ \bibnamefont {Bassett}},\ }\href@noop {} {\bibfield
  {journal} {\bibinfo  {journal} {Sci Rep}\ }\textbf {\bibinfo {volume} {6}},\
  \bibinfo {pages} {30770} (\bibinfo {year} {2016})}\BibitemShut {NoStop}%
\bibitem [{\citenamefont {Stiso}\ \emph {et~al.}(2019)\citenamefont {Stiso},
  \citenamefont {Khambhati}, \citenamefont {Menara}, \citenamefont {Kahn},
  \citenamefont {Stein}, \citenamefont {Das}, \citenamefont {Gorniak},
  \citenamefont {Tracy}, \citenamefont {Litt}, \citenamefont {Davis},
  \citenamefont {Pasqualetti}, \citenamefont {Lucas},\ and\ \citenamefont
  {Bassett}}]{stiso2019white}%
  \BibitemOpen
  \bibfield  {author} {\bibinfo {author} {\bibfnamefont {J.}~\bibnamefont
  {Stiso}}, \bibinfo {author} {\bibfnamefont {A.~N.}\ \bibnamefont
  {Khambhati}}, \bibinfo {author} {\bibfnamefont {T.}~\bibnamefont {Menara}},
  \bibinfo {author} {\bibfnamefont {A.~E.}\ \bibnamefont {Kahn}}, \bibinfo
  {author} {\bibfnamefont {J.~M.}\ \bibnamefont {Stein}}, \bibinfo {author}
  {\bibfnamefont {S.~R.}\ \bibnamefont {Das}}, \bibinfo {author} {\bibfnamefont
  {R.}~\bibnamefont {Gorniak}}, \bibinfo {author} {\bibfnamefont
  {J.}~\bibnamefont {Tracy}}, \bibinfo {author} {\bibfnamefont
  {B.}~\bibnamefont {Litt}}, \bibinfo {author} {\bibfnamefont {K.~A.}\
  \bibnamefont {Davis}}, \bibinfo {author} {\bibfnamefont {F.}~\bibnamefont
  {Pasqualetti}}, \bibinfo {author} {\bibfnamefont {T.~H.}\ \bibnamefont
  {Lucas}}, \ and\ \bibinfo {author} {\bibfnamefont {D.~S.}\ \bibnamefont
  {Bassett}},\ }\href@noop {} {\bibfield  {journal} {\bibinfo  {journal} {Cell
  Rep}\ }\textbf {\bibinfo {volume} {28}},\ \bibinfo {pages} {2554} (\bibinfo
  {year} {2019})}\BibitemShut {NoStop}%
\bibitem [{\citenamefont {Cui}\ \emph {et~al.}(2019)\citenamefont {Cui},
  \citenamefont {Li}, \citenamefont {Xia}, \citenamefont {Larsen},
  \citenamefont {Adebimpe}, \citenamefont {Baum}, \citenamefont {Cieslak},
  \citenamefont {Gur}, \citenamefont {Gur}, \citenamefont {Moore},
  \citenamefont {Oathes}, \citenamefont {Alexander-Bloch}, \citenamefont
  {Raznahan}, \citenamefont {Roalf}, \citenamefont {Shinohara}, \citenamefont
  {Wolf}, \citenamefont {Davatzikos}, \citenamefont {Bassett}, \citenamefont
  {Fair}, \citenamefont {Fan},\ and\ \citenamefont
  {Satterthwaite}}]{cui2019individual}%
  \BibitemOpen
  \bibfield  {author} {\bibinfo {author} {\bibfnamefont {Z.}~\bibnamefont
  {Cui}}, \bibinfo {author} {\bibfnamefont {H.}~\bibnamefont {Li}}, \bibinfo
  {author} {\bibfnamefont {C.~H.}\ \bibnamefont {Xia}}, \bibinfo {author}
  {\bibfnamefont {B.}~\bibnamefont {Larsen}}, \bibinfo {author} {\bibfnamefont
  {A.}~\bibnamefont {Adebimpe}}, \bibinfo {author} {\bibfnamefont {G.~L.}\
  \bibnamefont {Baum}}, \bibinfo {author} {\bibfnamefont {M.}~\bibnamefont
  {Cieslak}}, \bibinfo {author} {\bibfnamefont {R.~E.}\ \bibnamefont {Gur}},
  \bibinfo {author} {\bibfnamefont {R.~C.}\ \bibnamefont {Gur}}, \bibinfo
  {author} {\bibfnamefont {T.~M.}\ \bibnamefont {Moore}}, \bibinfo {author}
  {\bibfnamefont {D.~J.}\ \bibnamefont {Oathes}}, \bibinfo {author}
  {\bibfnamefont {A.}~\bibnamefont {Alexander-Bloch}}, \bibinfo {author}
  {\bibfnamefont {A.}~\bibnamefont {Raznahan}}, \bibinfo {author}
  {\bibfnamefont {D.~R.}\ \bibnamefont {Roalf}}, \bibinfo {author}
  {\bibfnamefont {R.~T.}\ \bibnamefont {Shinohara}}, \bibinfo {author}
  {\bibfnamefont {D.~H.}\ \bibnamefont {Wolf}}, \bibinfo {author}
  {\bibfnamefont {C.}~\bibnamefont {Davatzikos}}, \bibinfo {author}
  {\bibfnamefont {D.~S.}\ \bibnamefont {Bassett}}, \bibinfo {author}
  {\bibfnamefont {D.~A.}\ \bibnamefont {Fair}}, \bibinfo {author}
  {\bibfnamefont {Y.}~\bibnamefont {Fan}}, \ and\ \bibinfo {author}
  {\bibfnamefont {T.~D.}\ \bibnamefont {Satterthwaite}},\ }\href@noop {}
  {\bibfield  {journal} {\bibinfo  {journal} {bioRxiv}\ }\textbf {\bibinfo
  {volume} {694489}} (\bibinfo {year} {2019})}\BibitemShut {NoStop}%
\bibitem [{\citenamefont {Roalf}\ \emph {et~al.}(2016)\citenamefont {Roalf},
  \citenamefont {Quarmley}, \citenamefont {Elliott}, \citenamefont
  {Satterthwaite}, \citenamefont {Vandekar}, \citenamefont {Ruparel},
  \citenamefont {Gennatas}, \citenamefont {Calkins}, \citenamefont {Moore},
  \citenamefont {Hopson}, \citenamefont {Prabhakaran}, \citenamefont {Jackson},
  \citenamefont {Verma}, \citenamefont {Hakonarson}, \citenamefont {Gur},\ and\
  \citenamefont {Gur}}]{Roalf2016903}%
  \BibitemOpen
  \bibfield  {author} {\bibinfo {author} {\bibfnamefont {D.~R.}\ \bibnamefont
  {Roalf}}, \bibinfo {author} {\bibfnamefont {M.}~\bibnamefont {Quarmley}},
  \bibinfo {author} {\bibfnamefont {M.~A.}\ \bibnamefont {Elliott}}, \bibinfo
  {author} {\bibfnamefont {T.~D.}\ \bibnamefont {Satterthwaite}}, \bibinfo
  {author} {\bibfnamefont {S.~N.}\ \bibnamefont {Vandekar}}, \bibinfo {author}
  {\bibfnamefont {K.}~\bibnamefont {Ruparel}}, \bibinfo {author} {\bibfnamefont
  {E.~D.}\ \bibnamefont {Gennatas}}, \bibinfo {author} {\bibfnamefont {M.~E.}\
  \bibnamefont {Calkins}}, \bibinfo {author} {\bibfnamefont {T.~M.}\
  \bibnamefont {Moore}}, \bibinfo {author} {\bibfnamefont {R.}~\bibnamefont
  {Hopson}}, \bibinfo {author} {\bibfnamefont {K.}~\bibnamefont {Prabhakaran}},
  \bibinfo {author} {\bibfnamefont {C.~T.}\ \bibnamefont {Jackson}}, \bibinfo
  {author} {\bibfnamefont {R.}~\bibnamefont {Verma}}, \bibinfo {author}
  {\bibfnamefont {H.}~\bibnamefont {Hakonarson}}, \bibinfo {author}
  {\bibfnamefont {R.~C.}\ \bibnamefont {Gur}}, \ and\ \bibinfo {author}
  {\bibfnamefont {R.~E.}\ \bibnamefont {Gur}},\ }\href {\doibase
  http://dx.doi.org/10.1016/j.neuroimage.2015.10.068} {\bibfield  {journal}
  {\bibinfo  {journal} {NeuroImage}\ }\textbf {\bibinfo {volume} {125}},\
  \bibinfo {pages} {903 } (\bibinfo {year} {2016})}\BibitemShut {NoStop}%
\bibitem [{\citenamefont {Vandekar}\ \emph {et~al.}(2015)\citenamefont
  {Vandekar}, \citenamefont {Shinohara}, \citenamefont {Raznahan},
  \citenamefont {Roalf}, \citenamefont {Ross}, \citenamefont {DeLeo},
  \citenamefont {Ruparel}, \citenamefont {Verma}, \citenamefont {Wolf},
  \citenamefont {Gur}, \citenamefont {Gur},\ and\ \citenamefont
  {Satterthwaite}}]{Vandekar14012015}%
  \BibitemOpen
  \bibfield  {author} {\bibinfo {author} {\bibfnamefont {S.~N.}\ \bibnamefont
  {Vandekar}}, \bibinfo {author} {\bibfnamefont {R.~T.}\ \bibnamefont
  {Shinohara}}, \bibinfo {author} {\bibfnamefont {A.}~\bibnamefont {Raznahan}},
  \bibinfo {author} {\bibfnamefont {D.~R.}\ \bibnamefont {Roalf}}, \bibinfo
  {author} {\bibfnamefont {M.}~\bibnamefont {Ross}}, \bibinfo {author}
  {\bibfnamefont {N.}~\bibnamefont {DeLeo}}, \bibinfo {author} {\bibfnamefont
  {K.}~\bibnamefont {Ruparel}}, \bibinfo {author} {\bibfnamefont
  {R.}~\bibnamefont {Verma}}, \bibinfo {author} {\bibfnamefont {D.~H.}\
  \bibnamefont {Wolf}}, \bibinfo {author} {\bibfnamefont {R.~C.}\ \bibnamefont
  {Gur}}, \bibinfo {author} {\bibfnamefont {R.~E.}\ \bibnamefont {Gur}}, \ and\
  \bibinfo {author} {\bibfnamefont {T.~D.}\ \bibnamefont {Satterthwaite}},\
  }\href {\doibase 10.1523/JNEUROSCI.3628-14.2015} {\bibfield  {journal}
  {\bibinfo  {journal} {The Journal of Neuroscience}\ }\textbf {\bibinfo
  {volume} {35}},\ \bibinfo {pages} {599} (\bibinfo {year} {2015})}\BibitemShut
  {NoStop}%
\bibitem [{\citenamefont {Daducci}\ \emph {et~al.}(2012)\citenamefont
  {Daducci}, \citenamefont {Gerhard}, \citenamefont {Griffa}, \citenamefont
  {Lemkaddem}, \citenamefont {Cammoun}, \citenamefont {Gigandet}, \citenamefont
  {Meuli}, \citenamefont {Hagmann},\ and\ \citenamefont
  {Thiran}}]{10.1371/journal.pone.0048121}%
  \BibitemOpen
  \bibfield  {author} {\bibinfo {author} {\bibfnamefont {A.}~\bibnamefont
  {Daducci}}, \bibinfo {author} {\bibfnamefont {S.}~\bibnamefont {Gerhard}},
  \bibinfo {author} {\bibfnamefont {A.}~\bibnamefont {Griffa}}, \bibinfo
  {author} {\bibfnamefont {A.}~\bibnamefont {Lemkaddem}}, \bibinfo {author}
  {\bibfnamefont {L.}~\bibnamefont {Cammoun}}, \bibinfo {author} {\bibfnamefont
  {X.}~\bibnamefont {Gigandet}}, \bibinfo {author} {\bibfnamefont
  {R.}~\bibnamefont {Meuli}}, \bibinfo {author} {\bibfnamefont
  {P.}~\bibnamefont {Hagmann}}, \ and\ \bibinfo {author} {\bibfnamefont
  {J.-P.}\ \bibnamefont {Thiran}},\ }\href {\doibase
  10.1371/journal.pone.0048121} {\bibfield  {journal} {\bibinfo  {journal}
  {PLoS ONE}\ }\textbf {\bibinfo {volume} {7}},\ \bibinfo {pages} {1} (\bibinfo
  {year} {2012})}\BibitemShut {NoStop}%
\bibitem [{\citenamefont {Watts}\ and\ \citenamefont
  {Strogatz}(1998)}]{watts1998}%
  \BibitemOpen
  \bibfield  {author} {\bibinfo {author} {\bibfnamefont {D.~J.}\ \bibnamefont
  {Watts}}\ and\ \bibinfo {author} {\bibfnamefont {S.~H.}\ \bibnamefont
  {Strogatz}},\ }\href {\doibase 10.1038/30918} {\bibfield  {journal} {\bibinfo
   {journal} {Nature}\ }\textbf {\bibinfo {volume} {393}},\ \bibinfo {pages}
  {440} (\bibinfo {year} {1998})}\BibitemShut {NoStop}%
\bibitem [{\citenamefont {Rubinov}\ and\ \citenamefont
  {Sporns}(2010)}]{rubinov2010}%
  \BibitemOpen
  \bibfield  {author} {\bibinfo {author} {\bibfnamefont {M.}~\bibnamefont
  {Rubinov}}\ and\ \bibinfo {author} {\bibfnamefont {O.}~\bibnamefont
  {Sporns}},\ }\href@noop {} {\bibfield  {journal} {\bibinfo  {journal}
  {Neuroimage}\ }\textbf {\bibinfo {volume} {52}},\ \bibinfo {pages} {1059}
  (\bibinfo {year} {2010})}\BibitemShut {NoStop}%
\bibitem [{\citenamefont {Newman}(2004)}]{newman2004}%
  \BibitemOpen
  \bibfield  {author} {\bibinfo {author} {\bibfnamefont {M.~E.~J.}\
  \bibnamefont {Newman}},\ }\href {\doibase 10.1103/PhysRevE.69.066133}
  {\bibfield  {journal} {\bibinfo  {journal} {Phys. Rev. E}\ }\textbf {\bibinfo
  {volume} {69}},\ \bibinfo {pages} {066133} (\bibinfo {year}
  {2004})}\BibitemShut {NoStop}%
\bibitem [{\citenamefont {Borgatti}\ and\ \citenamefont
  {Everett}(2000)}]{borgatti2000}%
  \BibitemOpen
  \bibfield  {author} {\bibinfo {author} {\bibfnamefont {S.~P.}\ \bibnamefont
  {Borgatti}}\ and\ \bibinfo {author} {\bibfnamefont {M.~G.}\ \bibnamefont
  {Everett}},\ }\href {\doibase https://doi.org/10.1016/S0378-8733(99)00019-2}
  {\bibfield  {journal} {\bibinfo  {journal} {Social Networks}\ }\textbf
  {\bibinfo {volume} {21}},\ \bibinfo {pages} {375 } (\bibinfo {year}
  {2000})}\BibitemShut {NoStop}%
\end{thebibliography}%

\end{document}